\documentclass[preprint,prx,onecolumn,superscriptaddress,floatfix]{revtex4-1}

\usepackage{amsmath,amssymb}

\usepackage{changepage}

\usepackage[utf8x]{inputenc}

\usepackage{textcomp,marvosym}

\usepackage{nameref,hyperref}

\usepackage[right]{lineno}

\usepackage{microtype}
\DisableLigatures[f]{encoding = *, family = * }

\usepackage[table]{xcolor}

\usepackage{array}
\usepackage{soul}

\usepackage{amsthm}
\usepackage{mathtools} 
\usepackage{cases}
\usepackage{amsfonts}

\makeatletter
\renewcommand{\@biblabel}[1]{\quad#1.}
\makeatother

\begin{document}

\title{ Maximal information transmission is compatible with ultrasensitive biological pathways }
%\\
\author{Gabriele Micali}
%\thanks{Current Address:  Department of Environmental Microbiology, Eawag, D\"{u}bendorf, and Department of Environmental Systems Science, ETH Z\"{u}rich, Z\"{u}rich, Switzerland}
\affiliation{Department of Life Sciences, Imperial College, London, UK}
\affiliation{Centre for Integrative Systems Biology and Bioinformatics at Bioinformatics, Imperial College, London, UK}
\affiliation{Department of Environmental Microbiology, Eawag, D\"{u}bendorf, Switzerland}
\affiliation{Department of Environmental Systems Science, ETH Z\"{u}rich, Z\"{u}rich, Switzerland}
\author{Robert G. Endres}
\email{Corresponding Author: r.endres@imperial.ac.uk}
% Gabriele Micali\textsuperscript{1,2,\textcurrency} %\Yinyang},
% Robert G. Endres\textsuperscript{1,2,*} %\Yinyang},
\affiliation{Department of Life Sciences, Imperial College, London, UK}
\affiliation{Centre for Integrative Systems Biology and Bioinformatics at Bioinformatics, Imperial College, London, UK}

% Insert additional author notes using the symbols described below. Insert symbol callouts after author names as necessary.
% 
% Remove or comment out the author notes below if they aren't used.
%
% Primary Equal Contribution Note
%%\Yinyang These authors contributed equally to this work.

% Additional Equal Contribution Note
% Also use this double-dagger symbol for special authorship notes, such as senior authorship.
%%\ddag These authors also contributed equally to this work.

%\textcurrency Current Address:  Department of Environmental Microbiology, Eawag, D\"{u}bendorf, and Department of Environmental Systems Science, ETH Z\"{u}rich, Z\"{u}rich, Switzerland % change symbol to "\textcurrency a" if more than one current address note
% \textcurrency b Insert second current address 
% \textcurrency c Insert third current address

%%\dag Deceased

% Group/Consortium Author Note
%%\textpilcrow Membership list can be found in the Acknowledgments section.

% Use the asterisk to denote corresponding authorship and provide email address in note below.
%* r.endres@imperial.ac.uk

%\section*{Abstract}
\begin{abstract}
Cells are often considered input-output devices that maximize the transmission of information by converting extracellular stimuli (input) via signaling pathways (communication channel) to cell behavior (output). However, in biological systems outputs might feed back into inputs due to cell motility, and the biological channel can change by mutations during evolution. Here, we show that the conventional channel capacity obtained by optimizing the input distribution for a fixed channel may not reflect the global optimum. In a new approach we analytically identify both input distributions and input-output curves that optimally transmit information, given constraints from noise and the dynamic range of the channel. We find a universal optimal input distribution only depending on the input noise, and we generalize our formalism to multiple outputs (or inputs). Applying our formalism to \textit{Escherichia coli} chemotaxis, we find that its pathway is compatible with optimal information transmission despite the ultrasensitive rotary motors.
\end{abstract}

\maketitle

%\section*{Author summary}
%Information theory has long been applied in engineering applications for describing the transmission of information and its limits. However, applying such concepts to biological signaling pathways has been difficult: Pathways are not hard-wired as in computer chips but can change by evolution, and the final signaling output can affect the input as cells may move, making information flow a circular problem. Hence, whether biological organisms actually maximize information flow and how to test this theoretically have been open questions. Here, we propose a new approach by allowing not only the input (or output) to change, e.g. by changes in behavior, but also the pathway, e.g. by evolution. For an illustrative example we apply our approach to the \textit{Escherichia coli} chemosensory pathway, which allows these bacteria to swim up chemical gradients, e.g. to find nutrients. Although an individual motor used for swimming is highly sensitive and switch-like, and hence cannot transmit much more than a single bit of information, we find that this pathway nevertheless transmits information maximally due to multiple motors working together. Our framework is sufficiently general, allowing the testing of maximal information flow in other biological systems.

% \linenumbers

% Use "Eq" instead of "Equation" for equation citations.
\section*{Introduction}
Biological cells continuously process environmental cues, allowing them to make critical decisions quickly, e.g. whether to move or stay, whether to express a certain protein, or whether to divide \cite{Swain_msbREW}. These decisions are generally made based on the level of one or more key proteins, which are the internal representation of the extracellular stimulus. The higher the amount of environmental information encoded in the intracellular representation, the more reliable the response. In contrast, cell-external and internal noise may reduce the reliability. Hence, a biological system under evolutionary pressure is expected to evolve to optimally transmit information under biologically relevant constraints \cite{Vergassola2007infotaxis}, at least when information is a limiting factor \cite{Lander13,Taylor2007information,Rivoire_Leibler_11}. 

\begin{figure}
\includegraphics{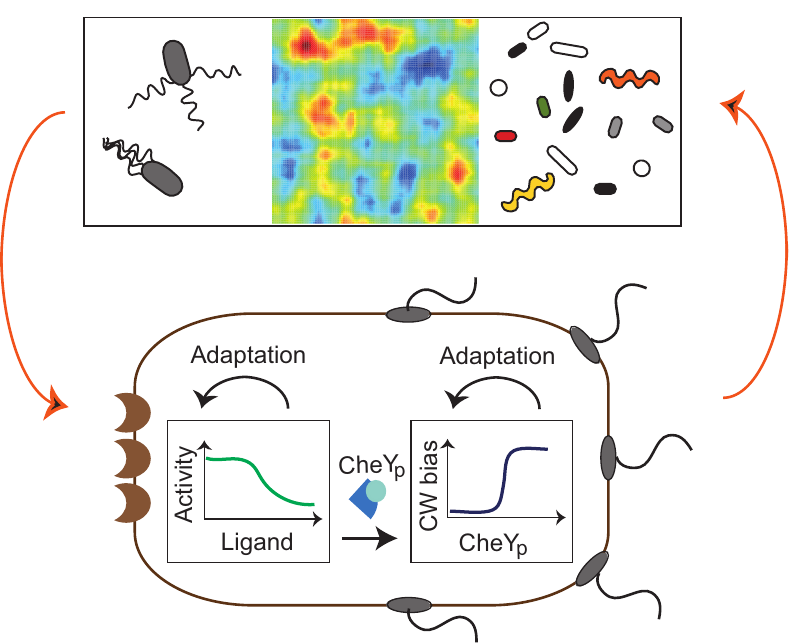}
 \caption{\textbf{Connection between environmental cues and chemotactic response.}  Chemotactic bacteria live in complex microenvironments in which input distributions of chemical concentrations are shaped by the swimming behavior (top left), chemical sources and sinks (top middle), and competition with other bacteria (top right). Inputs are processed by the cell-internal chemotaxis pathway, which can be viewed as an input-output device (bottom). Specifically, input-output curves are measured in experiments by dose-response curves with noise. The resulting final behavior feeds back into the environment. Evolution is assumed to select the best input-output curves for maximizing fitness. The chemotactic pathway is a two-component system, and for modeling purposes, is divided into two information-transmission channels: receptors sense external concentration of stimuli and their activity regulates the protein $\text{CheY}_\text{p}$ (receptor channel, bottom left). $\text{CheY}_\text{p}$ is the internal representation of the external stimulus and regulates motor switching (clockwise or counterclockwise rotation) and thus bacterial motility (straight swimming via a `run' or random reorientation via a `tumble'; motility channel, bottom right). Note that there is additional adaptation both at the receptors \cite{SouBerg02a} and the motors \cite{YuanBerg12Nat}.}\label{Fig1}
\end{figure}

To formalize this optimization problem, cells can be considered input-output devices, where stimuli of extracellular concentrations are the input, receptors (or the entire pathway) are the communication channel, and the intracellular concentration of a key protein (or the final behavior of the cell) is the output. 
%However, in biological systems with feedback the distinction between input, channel, and output is not always clear. 
In contrast to engineered physical systems, the distinction between input, channel, and output is not always clear in biological systems with feedback. Take for instance \textit{Escherichia coli} chemotaxis, a well-characterized pathway allowing bacteria to sense chemicals and to swim towards nutrients \cite{Berg00,EndresBook13}. The intracellular level of the phosphorylated protein ($\text{CheY}_\text{p}$) represents the extracellular concentration of a chemical and regulates the motors (clockwise or counterclockwise rotation) and hence motility (`run' or `tumble') (Fig. \ref{Fig1}). The swimming behavior clearly affects the input as cells change their location, making information flow a circular problem. Hence, the question emerges how to tackle such problems.

Shannon's mutual information is generally used to quantify information transmission, capturing the statistical (linear and nonlinear) dependencies between inputs and outputs \cite{Shannon}. 
Specifically, the mutual information describes the ability on average to reconstruct the input distribution after repeatedly measuring the output \cite{Shannon,SwainRevInfo14}. 
Often maximal mutual information is assumed, either to reflect biological function or because the mutual information cannot be calculated otherwise \cite{Tkavcik_11Rev, Tkacik_16Rev}. 
How should mutual information be maximized? Maximizing with respect to the (generally unknown) distribution of inputs leads to the channel capacity. 
Such an approach was, e.g., used to study transcriptional regulation in the developing fruit-fly embryo \cite{TkaCal08a, TkaCal08b}. While in this case, the mother organism may be able to tune its maternal factors to the optimal input distribution to match the `expectation' of the embryo, the channel capacity may not generally be a valid approach. 
Alternatively, it is possible to assume a fixed input distribution and to maximize the mutual information with respect to the input-output curve \cite{DetRamShr00, Micali14}. The underlying idea is that the input-output curve is adjusted by evolution to match the distribution of inputs. This approach might apply to organisms confined to certain environments, e.g. bacteria living in specific niches \cite{Micali14}. 
In addition to these two optimization procedures (and some attempts to combine them for specific types of input-output curves \cite{Tkavcik_09PRE, Walczak_10PRE, Tkavcik_12PRE}), the Fisher information from estimation theory may also be used to predict the input distribution \cite{bernardo1979reference,BrunelNadal}. 
Hence, due to multiple available approaches there is considerable uncertainty to what optimization procedure to use. 

Even the principle of maximal information transmission may be questioned. Recently, maximizing information transmission at the \textit{E. coli} receptors led to maximal drift of cells swimming up a chemical gradient and hence optimal chemotactic behavior \cite{Micali14}. Although a reasonable result, there are two potential problems when considering the whole pathway: Firstly, the dose-response curves of the motors are measured to have Hill coefficients up to 20 \cite{YuanBerg13}. Hence, such switch-like motors may only transmit about one bit of information, distinguishing only low and high levels of $\text{CheY}_\text{p}$. Secondly, the $\text{CheY}_\text{p}$ level, which maximizes the drift is not in the sensitive region of the motor \cite{Dufour14, Vergassola16, Micali17}. Both issues may lead one to suggest that high information transmission at the receptors is wasted downstream at the motors, and that an entirely different principle may guide cell behavior \cite{Skoge_Meir_Win_PRL11, ThomasP2016}. 

Here, we address two key questions: (1) How should the mutual information be calculated in a biological context, and (2) does the bacterial chemotaxis pathway maximize information transmission? 
Specifically, we reconcile the various optimization procedures, leading to a new way of maximizing the mutual information, particularly useful for biological systems. Assuming general external and internal noise, and a fixed range of sensitivity (as any biological or physical system is necessarily limited), we analytically derive both the optimal input distribution \textit{and} input-output curve. Unlike previous approaches \cite{Tkavcik_09PRE, Walczak_10PRE, Tkavcik_12PRE}, this general solution does not assume specific input-output curves, such as Hill functions. Surprisingly, we find an universal optimal input distribution, only dependent on the input noise. Furthermore, {numerically and with the help of simulations,} we were able to extend our formalism to multiple outputs (or inputs), greatly extending the applicability of our formalism to biological systems.  
As an illustrative example, we focus on the \textit{E. coli} chemotaxis pathway using previously estimated noise and measured dose-response curves at the receptors and the motors. By deriving analytical results for multiple output motors, we show that although the optimal response is not a Hill function, the measured Hill coefficients naturally emerge from our optimal prediction. Overall our results confirm the idea of maximal information transmission in \textit{E. coli} chemotaxis, and prove that maximal information transmission at the receptors is critical for the whole pathway despite the ultrasteep motor dose-response curves.

% Results and Discussion can be combined.
\section*{Results}

\subsection*{Maximizing mutual information: comparison of different approaches}

Information transmission between an input, $X$, and an output, $Y$, 
is often quantified by the mutual information, which is a measure of statistical dependency and reflects the average ability for inferring the input after measurements of the output (\cite{Shannon, Tkavcik_11Rev, Tkacik_16Rev, SwainRevInfo14, Nemenman_14Review, McMahon_14, Micali16} for extended reviews). %Among all the possible measures of statistical dependencies mutual information is considered highly relevant in biology (\cite{SwainRevInfo14, Nemenman_14Review, McMahon_14} for extended reviews). 
For continuous random variables, $X$ and $Y$, the mutual information is defined by
\begin{equation}
\label{eq:MI}
\mathcal{I}[X,Y]:= \int \text{d}y \, \text{d}x \, p(y|x) p(x) \text{log}_2 \left[ \frac{p(y|x)}{p(y)} \right] ,
\end{equation} 
where $p(y|x)$ is the conditional probability of observing $Y=y$ at given $X=x$, encoding the input-output curve and noise. Quantities $p(x)$ and $p(y)$ represent the input and output distributions, respectively, which are mathematically connected by the conservation of probability $p(x) \text{d}x = p(y) \text{d}y$, valid in the small-noise limit. %$G=\frac{\partial \bar{y}}{\partial x}$
Considering small Gaussian noise with conditional probability, $p(y|x)=\exp\left[-(y-\bar{y})^2/(2 \sigma_\text{T}^2) \right] /\sqrt{2 \pi \sigma_\text{T}^2}$, with mean $\bar{y}(x)$ and standard deviation $\sigma_\text{T}(x)$, Eq. \eqref{eq:MI} becomes 
\begin{equation}
\label{eq:MI2}
\mathcal{I}[X,Y]= - \int_{x_\text{on}}^{x_\text{off}} \text{d}x \, p(x) \, \text{log}_2 \left[ \frac{\sqrt{2 \pi e} \ \sigma_\text{T}(x) }{\bar{y}'(x)} p(x) \right] , 
\end{equation} 
where $x_\text{on}$ and $x_\text{off}$ set the sensitive region, i.e. the dynamic range of inputs \cite{DetRamShr00,Micali14} (similar equations appear in \cite{TkaCal08a,TkaCal08b,BrunelNadal}). In addition to $p(x)$, Eq. \eqref{eq:MI2} depends on the gain $\bar{y}'(x)$, i.e. the first derivative of the input-output curve, $\bar{y}'(x)$, and total noise $\sigma_\text{T}(x)$.

To understand if biological systems maximize information flow, we need to maximise the mutual information and derive general principles or compare with data. 
To maximise the mutual information, the channel capacity is often considered, i.e. the mutual information maximized with respect to the input distribution \cite{TkaCal08a, TkaCal08b}. 
Alternatively, the mutual information can be maximized with respect to the input-output curve assuming a fixed input distribution \cite{DetRamShr00, Micali14}. The former method is an attempt to deal with the often unknown input distribution, while the latter is based on the idea that the biological channel can be modified by evolution. %Can the different methods be unified? 
Additionally, the two approaches were combined for specific input-output Hill and Hill-like functions \cite{Tkavcik_09PRE, Walczak_10PRE, Tkavcik_12PRE}. However, is there a general way to unify the different methods, without making assumptions about functional form of the input-output functions?

Formally, we maximise the mutual information, Eq. \eqref{eq:MI2}, with respect to $p(x)$ and $\bar{y}(x)$ by writing
\begin{align}
\label{eq:sist1}
 \max_{p(x)} \mathcal{I}[X,Y]  \ \ &\rightarrow \ \ \ \frac{\partial \mathcal{L}}{\partial p} =0,\\
\label{eq:sist2}
 \max_{\bar{y}(x)} \mathcal{I}[X,Y] \ \  &\rightarrow \ \ \ \frac{\partial \mathcal{L}}{\partial \bar{y}} - \frac{\text{d} }{\text{d} x} \frac{\partial \mathcal{L}}{\partial \bar{y}'} = 0 ,
\end{align}
where the right-hand side of Eqs. \eqref{eq:sist1} and \eqref{eq:sist2} are Euler-Lagrange equations from calculus of variations with Lagrangian $\mathcal{L}(x, p, \bar{y}, \bar{y}')= p \, \text{log}_2 \left[ \frac{\sqrt{2 \pi e} \sigma_\text{T} }{\bar{y}'} p \right]$, i.e. the integrand of Eq. \eqref{eq:MI2}. 
Equation \eqref{eq:sist1} represents the channel capacity applied to a Gaussian channel (i.e. Gaussian conditional probability $p(y|x)$, see \cite{TkaCal08a,TkaCal08b} for examples in gene regulation). In contrast, Eq. \eqref{eq:sist2} is used to obtain the optimal input-output curve for a given input distribution (see \cite{DetRamShr00,Micali14} for examples in sensory systems). For completeness, we provide the solutions of the individual maximizations of Eqs. \eqref{eq:sist1} and \eqref{eq:sist2} in \textit{SI Text} Sec. 1.1-1.2, with a discussion of the sensitive region in Sec 1.8. 

\begin{figure}
\includegraphics{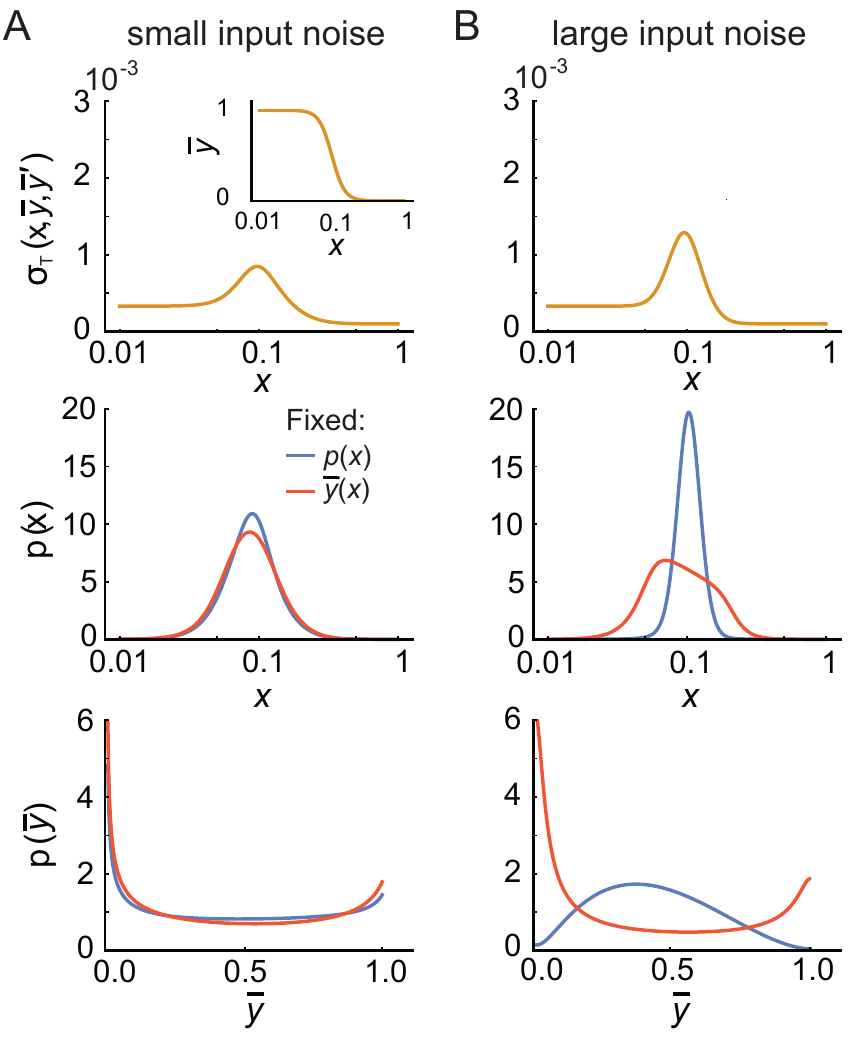}
 \caption{\textbf{Conventional ways of maximizing mutual information.} 
Comparison of the channel capacity, i.e. solving the maximization problem Eq. \eqref{eq:sist1} for $p(x)$ using a fixed $\bar{y}(x)$ and noise (red), and maximization with respect to input-output curve, i.e. solving the maximization problem Eq. \eqref{eq:sist2} for $\bar{y}(x)$ using a fixed $p(x)$ and noise (blue). (For specific examples of the solutions of Eq. \eqref{eq:sist1} and \eqref{eq:sist2} and a discussion of the bimodality of the output distribution, see \textit{SI Text} Sec. 1). 
In both cases, the noise is provided as a function of the input distribution and input-output curve (top row), with the input-output curve $\bar{y}(x)=\left[ 1+ (x/k_d)^{n_H}\right]^{-1}$ assumed a Hill function for simplicity, with Hill coefficient $n_H$ and threshold $k_d$ (inset). (A) For small input noise, the two approaches converge, i.e. the red and the blue input (middle left) and output (bottom left) distributions match. (B) For large input noise, the two approaches predict different input (middle right) and output (bottom right) distributions. 
 Using Eq. \eqref{eq:noise} for the noise, the input noise is $\sigma^2_x = \alpha_1 x$, while the output noise $\sigma^2_y = \alpha_2 \bar{y}(1-\bar{y}) + \alpha_3 \bar{y} + \alpha_4$ has three different contributions with $\bar{y}$ the input-output curve. The parameters are chosen to provide an overall similar level of noise, given by $\alpha_1 = 10^{-8}$, $\alpha_2 = 2 \cdot 10^{-6}$, $\alpha_3 = 10^{-7}$, $\alpha_4 = 10^{-8}$ (panel A) and $\alpha_1 = 10^{-7}$, $\alpha_2 = 10^{-8}$, $\alpha_3 = 10^{-7}$, $\alpha_4 = 10^{-8}$ (panel B). For the red model, $n_H=5$ and $k_d=0.1$. For the blue model, the input distribution is fixed by normalizing the derivative of a Hill function with $n_H=5$ and $k_d=0.1$, with the input-output curve free to change according to the maximization. The sensitive region is set by $x_\textit{on}=0$ and $x_\textit{off}= +\infty$.
 }\label{Fig2}
\end{figure}

When the noise is uniform ($\sigma_T$ constant) Eqs. \eqref{eq:sist1} and \eqref{eq:sist2} coincide. Specifically, $\frac{\partial \mathcal{L}}{\partial \bar{y}} = 0$ in Eq. \eqref{eq:sist2} so that $\bar{y}$ is a cyclic variable. As a result, $\frac{\text{d} } {\text{d} x} \frac{\partial p } {\partial \bar{y}'} = 0$ so that $p/\bar{y}' =$const and hence is conserved (not in time but in input space), following the Emmy Noether theorem. 
In this case, maximizing the mutual information leads to a simple matching relationship  ($\bar{y}' \propto p$), so that the input-output curve is the cumulative integral of the input distribution (see \textit{S1 text}, Sec. 1.1 and Fig. S1) \cite{Lau81}. However, in general when both input and output noise matter the noise is a function of the input and input-output curve, given by $\sigma_\text{T}=\sigma_\text{T}(x,\bar{y},\bar{y}')$. Assuming independent cell-external and internal noise, we consider 
\begin{equation}
\label{eq:noise}
\sigma_\text{T}(x,\bar{y},\bar{y}')= \sqrt{\sigma_x^2 \, \bar{y}'^2 + \sigma_y^2},
\end{equation}
which follows from error propagation. Specifically, $\sigma_x$ is the input noise depending on $x$ only, amplified by the gain $\bar{y}'(x)$, and $\sigma_y$ is the output noise depending on $\bar{y}(x)$ only. In case of negligible input ($\sigma_\text{x}\approx0$) or output ($\sigma_\text{y}\approx0$) noise, Eq. \eqref{eq:sist1} again converges to Eq. \eqref{eq:sist2} and the system can be solved for any input-output curve $\bar{y}(x)$ (see \textit{S1 text}, Sec. 1.2-1.3). As a result, the predicted input and output distributions from the two optimization approaches become identical (Fig. \ref{Fig2}A). However, in general the two equations differ and the resulting optimal input and output distributions are very different in the two approaches (Fig. \ref{Fig2}B). 
In particular, the output distributions can be uni- or bimodal with details described in \textit{SI Text}, Sec. 1.3. 

It is worth noting that, in Bayesian statistics the Fisher information is linked to the channel capacity \cite{bernardo1979reference,BrunelNadal}. A key problem in Bayesian statistics is choosing a prior distribution for a given stochastic process (i.e. $p(y|x)$) \cite{JeffreysP}. The idea of having a prior which does not affect the posterior distribution (i.e. $p(x|y)$) is linked to maximal mutual information, given by the average Kullback-Leibler divergence between prior and posterior distributions. This prior distribution is called the reference prior \cite{bernardo1979reference}, given by $p(x) \propto \sqrt{\mathcal{F} (x)}$  
%. To find the reference prior statisticians use the channel capacity, Eq. \eqref{eq:sist1}, which is linked to the Fisher information ($\mathcal{F}$), 
with Fisher information $\mathcal{F}(x)=\displaystyle\int dy \, p(y|x) \left( \partial \log \left(p(y|x)\right) / \partial x  \right)^2$. 
As shown in Ref.~\cite{BrunelNadal} the Fisher information is the result of maximizing the equivalent of Eq. \eqref{eq:MI2} for a general (not necessarily Gaussian) conditional probability distribution (see \textit{S1 Text}, Sec. 2.). Hence, the channel capacity and the approach based on the Fisher information are equivalent.

\subsection*{Maximizing mutual information: a new approach}

The difference between Eqs. \eqref{eq:sist1} and \eqref{eq:sist2} is that Eq. \eqref{eq:sist1} assumes a fixed input-output curve and a variable input distribution, while Eq. \eqref{eq:sist2} assumes a fixed input distribution and a variable input-output curve. There might be situations in which one approach is more appropriate than the other but in a general biological context the two are intrinsically connected (Fig. \ref{Fig1}). From a mathematical point of view, Eqs. \eqref{eq:sist1} and \eqref{eq:sist2} can be combined and solved together, i.e. $\mathcal{I}[X,Y]$ maximized with respect to both $p(x)$ and $\bar{y}(x)$. Similar numerical double optimizations are common in rate distortion theory using, e.g., the Blahut algorithm \cite{Blahut72,Taylor2007information}.

In what follows, we provide the analytical solution for $p(x)$ and $\bar{y}(x)$ by solving Eqs. \eqref{eq:sist1}-\eqref{eq:sist2} together. We assume a fixed dynamical range of inputs set by  $x_\text{on}$ and $x_\text{off}$ (given by the receptor sensitivity), leading in return to a fixed dynamical range of outputs from $y(x_\text{on})=1$ to $y(x_\text{off})=0$. We consider Eq. \eqref{eq:MI2} with noise given by Eq. \eqref{eq:noise}. After a first integration, we obtain the formal solution
\begin{subnumcases}{}
\label{eq:dLdpSol}
p= \frac{\bar{y}'}{Z \sigma_\text{T}}, \label{eq:solSy1} \\
\sigma_x^2 \, \bar{y}'^2 = Q \sigma_y^2, \label{eq:solSy2}
\end{subnumcases}
where $Z$ and $Q$ are two constants set by normalization and boundary conditions, respectively (see \textit{Materials and Methods} and \textit{S1 Text}, Sec. 1.7). 

Equation \eqref{eq:solSy1} for the input distribution extends the matching relationship found in \cite{Lau81} to nonuniform noise. In the latter case the optimal input distribution weighs certain inputs more than uncertain inputs \cite{TkaCal08b,BrunelNadal,KomorowskiStumpf}. Equation \eqref{eq:solSy2} determines the input-output curve which maximizes the mutual information given the noise. A solution of the system of equations exists if the transmitted input noise can be expressed in terms of the output noise or vice versa (see \textit{S1 Text}, Sec. 1.7). While such a solution may seem very specific, it is certainly plausible, given enough time, that evolution eventually finds it. 

\begin{figure}
\includegraphics[width=13cm]{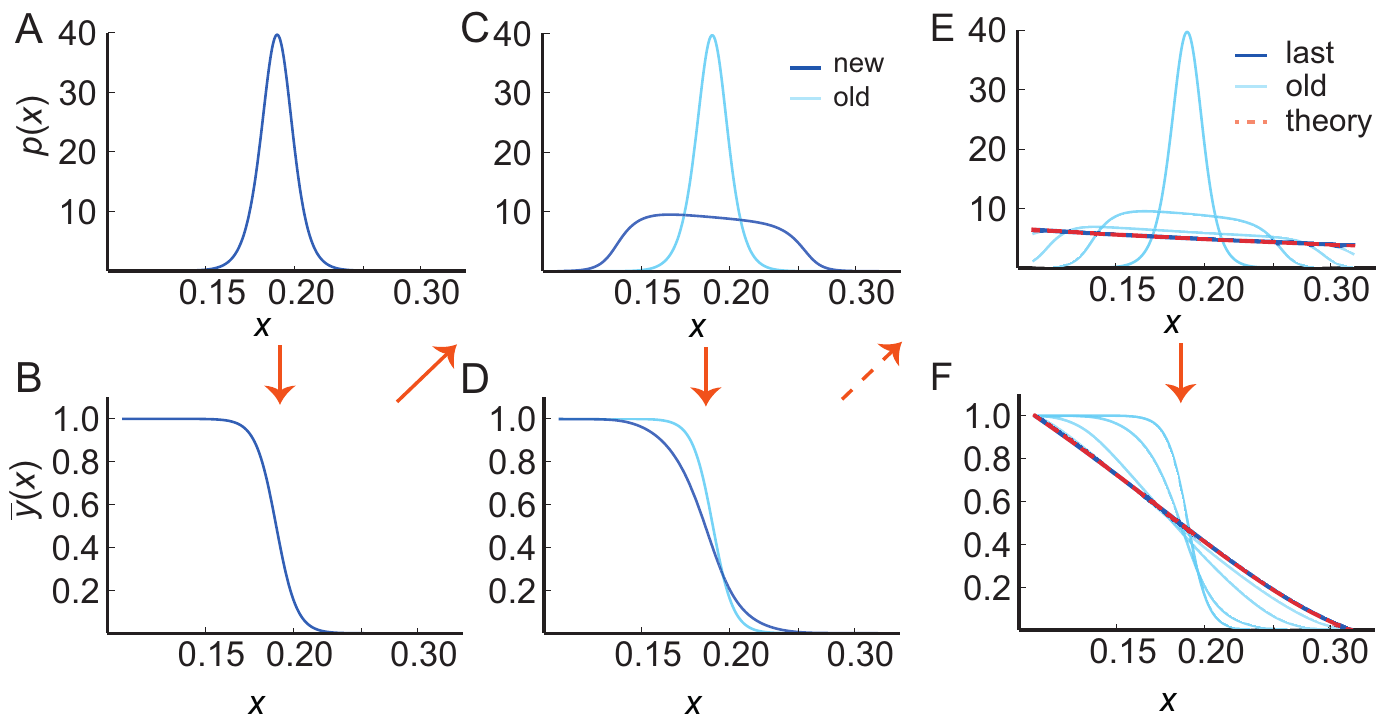}
 \caption{\textbf{Adaptive evolutionary algorithm approaches analytical result.} (A) Bacteria are assumed to live in a given environment and experience a given distribution of inputs. (B) Assuming that maximal information transmission enhances the chance to survive, evolution selects the mutations and hence the phenotype with optimal input-output curve. (C) At given input-output curve, the optimal input distribution generally (dark blue) differs from the initial input distribution (light blue) in (A). A change in behavior (and hence a change in the inputs) may provide an increase in information transmission (see Discussion section for more details). (D) The input-output curve changes again to maximise information transmission for the new set of stimuli.  (E,F) This iterative cycle continues and eventually converges to the solution given by Eqs. \eqref{eq:solSy1}-\eqref{eq:solSy2} (dashed red line). 
 Parameters: $\alpha_{1-3}=10^{-7}$, $\alpha_4=10^{-8}$, $x_{\text{on}}=0.115$, $x_{\text{off}}=0.323$.} 
 \label{Fig3}
\end{figure}

How may evolution find the solution? To mimic evolution, we envision an adaptive algorithm, allowing the pathway to iteratively reach optimal information transmission. Given an environment and hence a distribution of inputs, $p_1(x)$ (Fig. \ref{Fig3}A), evolution selects the optimal internal input-output curve, $y_1(x)$ (Fig. \ref{Fig3}B). However, the distribution of inputs is susceptible to changes, which might be caused by a change of the organism's behavior, even in the same environment.
The new input distribution, $p_2(x)$, may again lead to an increase in information transmission at fixed input-output curve, $y_1(x)$ (Fig. \ref{Fig3}C). Subsequently, evolution will select a new input-output curve, $y_2(x)$, which enhances information transmission at a fixed input distribution, $p_2(x)$ (Fig. \ref{Fig3}D). 
This cycle is repeated many times. If the optimal configuration is achievable and information transmission is a proxy for fitness, we expect that the solution of Eqs. \eqref{eq:solSy1} and \eqref{eq:solSy2} naturally emerges in the pathway. This is indeed the case for the examples studied here (see Fig. \ref{Fig3}E,F).

\subsection*{Information transmission at \textit{E. coli} chemoreceptors}

To apply our new approach, we use the chemotaxis pathway of \textit{E. coli} as an explicit example, since it is relatively simple and well characterized in its molecular components \cite{Berg00}. 
Briefly, chemoattractant (ligand) binding turns receptors off, inhibits the kinase CheA, and hence reduces the phospho-transfer from $\text{CheA}_\text{p}$ to CheY.  This leads to `runs' as only $\text{CheY}_\text{p}$ can bind the 6-8 motors to introduce `tumbling'. There is also an adaptation mechanism, where addition of methyl groups to receptors compensates for increased attractant concentration by increasing the receptor activity and hence the CheY$_\text{p}$ level to induce cell tumbling. 
Removal of methyl group has the opposite effect \cite{KeyEndSko06,EndSouWin2008}. In order to study signaling in fixed adaptational states, the adaptation enzymes can be removed from the chromosome and the receptor expressed with specific, genetically engineered, receptor modification levels to mimic receptor methylation (see \textit{S1 Text}, Sec. 4.3) \cite{KeyEndSko06,EndSouWin2008}.

Specifically, we consider the instantaneous information transmission between the chemoattractant methylaspartate (MeAsp) as the input and the response regulator $\text{CheY}_\text{p}$ as the output. Hence, we consider the information transmitted by the initial (fast) response for a given adaptational state (which only changes slowly). (At a later time this response is removed by adaptation and hence is transient only.) Note, unlike Ref. \cite{TosttenWolde09} we do not assume small Gaussian inputs but natural stimuli drawn from broad, potentially asymmetric input distributions $p(x)$. %These input distributions match the inputs experienced by cells swimming in gradients of arbitrary steepness \cite{Micali14}. 
When the input distribution of cells simulated in gradients of different strength match the optimal information-theoretical input distribution for the same receptor modification levels, the drift velocity up the gradient is maximized and, hence, this leads to optimal chemotaxis \cite{Micali14}.
As this matching of input distributions occurs anywhere in the gradient, these initial responses describe chemotaxis in the whole gradient, and so implicitly include adaptation. Indeed, the predicted distributions of inputs are scale invariant (when normalized by the adapted concentration) and reproduce Weber's law and logarithmic sensing \cite{Micali14}. The latter can also be captured by the predictive mutual information \cite{tenWolde15prl}.

The functional form of the noise is assumed to be known and derived from microscopic theory as in \cite{ClauEnd11BMC} (see \textit{S1 Text}, Secs. 3 and 4.2 for noise estimation and sensitivity to noise parameters, respectively). In short, the input variance is considered to be proportional to the input strength, $\sigma_x^2 = \alpha_1 x$, with $x$ in units of the ligand concentration the cell is adapted to. Furthermore, $\alpha_1 \propto (D N \tau)^{-1}$ is given by the Berg-and-Purcell limit \cite{BergPur77}, where $D$ is the diffusion constant of the ligand molecules, $N$ is the number of receptors acting cooperatively in a cluster, and $\tau$ is the averaging time, assuming a spherical cell \cite{BergPur77}. The output noise has three contributions: signaling noise, switching noise due to \textit{on}/\textit{off} changes of the receptor state, and a constant background noise, leading to $\sigma_y^2=\alpha_2  \bar{y} (1-\bar{y}) +\alpha_3 \bar{y} + \alpha_4$ with phosphorylated $\bar{y}$ in units of the total CheY level, $Y_T$, an intrinsic dependence on the (unknown) input-output curve, and $\alpha_{2-4}$ additional parameters defined in \textit{S1 Text}, Sec. 3. Note these effective noise terms are time-averaged due to their dependence on chemical reactions based on finite rate constants \cite{ClauEnd11BMC}. 
Despite $\sigma^2_y$ being specific, this noise should apply to many receptor-signaling pathways, including other two-component pathways \cite{Laub15_rev}. 
%
%{\color{blue}GM: Here $x$ is having dimension of a concentration, we should probably say at one point, maybe when we switch to $c$ that it is $c/c^*$ with $c^*=1 \mu M$ to get concentration adimentional.}

\begin{figure}
\includegraphics{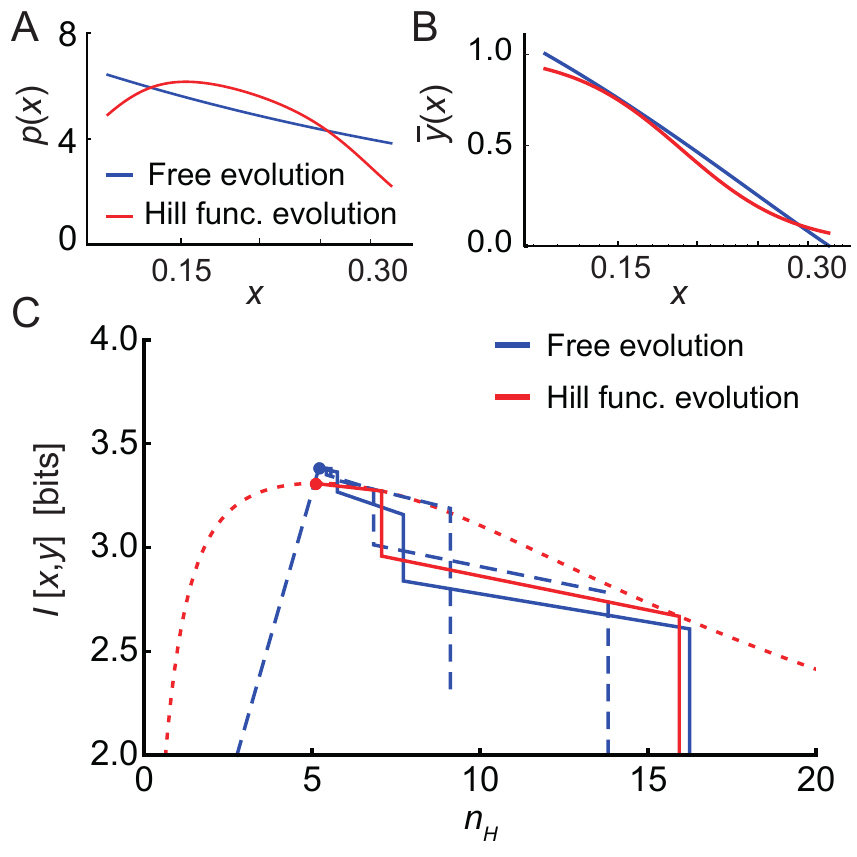}
 \caption{ \textbf{Robustness of the adaptive algorithm.} Comparison between our analytical result (blue lines) and the corresponding result when the input-output curve is constrained to a Hill function with adjustable Hill coefficient $n_H$ (red lines). (A,B) Input distributions (A) and the input-output curves (B). 
 (C) Comparison between the adaptive algorithm for the unconstrained (blue lines) and Hill-function constrained (red solid line) optimizations.
The adaptive algorithm moves the input-output curves to the optimal value $n_H\approx5$, steadily increasing the mutual information. 
The convergence towards the analytical solution is robust to different initial Hill coefficients (dashed blue lines). The optimal Hill coefficient is compatible with the corresponding experimental curve from \cite{EndSouWin2008} (see also Fig. S9). When constraining the input-output curve to a Hill function, the highest mutual information at a given Hill coefficient is shown by the red dashed line. Noise parameters: $\alpha_{1-3}=10^{-4}$, $\alpha_4=10^{-5}$, $x_{\text{on}}= 0.115$ and $x_{\text{off}}= 0.323$, (see \textit{S1 Text}, Sec. 3 for details of the noise).
}  
\label{Fig4}
\end{figure}

Using this noise, the explicit solution of Eq. \eqref{eq:solSy2}  is
\begin{align}
\bar{y}(x) &= \frac{\alpha_2 + \alpha_3 + \sqrt{\left( \alpha_2 + \alpha_3 \right)^2 + 4 \alpha_2 \alpha_4} \sin \left[ \sqrt{\alpha_2} \left( C + 2\sqrt{ \frac{Q x}{\alpha_1}} \right) \right]} {2 \alpha_2}, 
\label{eq:Sin} 
\end{align}
where $C$ and $Q$ are constants set by imposing fixed boundary conditions, $y(x_\text{off})=1$ and $y(x_\text{on})=0$ (Fig. \ref{Fig4}A). Note that $x$ appears with the prefactor $Q/\alpha_1$. Thus, $\alpha_1$ is set by the boundary conditions and it can be seen as the units of $x$. 
We consider two special cases: by simplifying the output noise for $\alpha_2=0$, we obtain  %recover the expansion terms, 
\begin{align}
\bar{y}(x) =  \frac{\alpha_3^2 C'^2 - 4 \alpha_4}{4 \alpha_3} + \alpha_3 C' \sqrt{\frac{Q x}{\alpha_1}} +\frac{\alpha_3 Q x}{\alpha_1},
\end{align}  
%where $x$ appears with a prefactor $Q/\alpha_1$. 
where the solution is again independent of $\alpha_1$ after imposing the boundary conditions. 
%Finally, 
For $\alpha_{2,3} =0$, we obtain
\begin{align}
\nonumber
 \bar{y}(x) &= 2 Q \sqrt{\frac{\alpha_4}{\alpha_1}} \sqrt{x} + C''  \\ 
 \label{eq:solsym}
  & = - \frac{ \sqrt{x} } { \sqrt{ x_\text{on}} - \sqrt{ x_\text{off}} } +  \frac{ \sqrt{ x_\text{on}}} { \sqrt{ x_\text{on}} - \sqrt{ x_\text{off}}}, 
\end{align}
which does not depend on both $\alpha_1$ and $\alpha_4$ once $Q$ and $C''$ are set by the boundary conditions. %(see \textit{Supporting Information Sec. 1.} for details on the noise). 
The optimal input distribution obtained by inserting Eq. \eqref{eq:solSy2} into \eqref{eq:dLdpSol} is $p =  (Z  \sqrt{(1+Q)/Q} \sigma_x)^{-1} $. In particular, for our choice of external noise and fixed sensitivity range, the input distribution converges to $p(x)=\left[ 2 (\sqrt{x_\text{on}} - \sqrt{x_\text{off}}) \sqrt{x} \right]^{-1} \sim 1/\sqrt{x}$ independently of $\alpha_{1-4}$ (see \textit{S1 Text}, Sec. 1.4.6 and Fig. S3). 
Importantly, this is a general result for the Berg-and-Purcell input noise, and hence should be valid for many signaling pathways. This result was previously found numerically \cite{Tkavcik_09PRE}, and such an input distribution of glutamine was suggested to optimize nitrogen sensing \cite{KomorowskiStumpf}.

Extracting the sensitive regime of \textit{E. coli} receptors for a fixed modification level (resembles receptor methylation level, see \textit{S1 Text}, Sec. 4.3) \cite{EndSouWin2008,EndWin06}, we test the convergence of the adaptive algorithm to the solution in Eq. \eqref{eq:Sin}. After a few iterative cycles the system indeed converges (Fig. \ref{Fig4}A), increasing the mutual information at each step (Fig. \ref{Fig4}C, blue line). 
This convergence to the analytical solution occurs when starting at different initial conditions, showing robustness of our algorithm. 
Note that in Fig. 4C the solution $\bar{y}(x)$ is fitted to Hill functions for convenience of presentation, allowing the mutual information to be plotted as a function of a single parameter (i.e. the Hill coefficient $n$). The optimal curve selects a Hill coefficient compatible to the experimental measurements from FRET data, at least for larger receptor modification levels (Fig. S9) \cite{EndSouWin2008}. 
Applying the adaptive algorithm instead to Hill-function constrained input-output curves produces the same optimal Hill coefficient $n$ albeit with a smaller mutual information (Fig. \ref{Fig4}C, red solid line, with Fig. \ref{Fig4}B comparing the corresponding optimal input distributions and optimal input-output curves). 
Note that for a fixed Hill equation the optimal mutual information is calculated directly using the input distribution from Eq. \eqref{eq:dLdpSol}, resulting in 
\begin{equation}
\mathcal{I}[x,y]=\text{log}_2 \left[ \frac{Z}{\sqrt{2 \pi \text{e}}} \right] \ \ \ \text{with} \ \ \ Z=\int_{x_\text{on}}^{x_\text{off}} dx \, \frac{\bar{y}'(x)}{\sigma_\text{T}(x)}, 
\end{equation}
as shown in Fig. \ref{Fig4}C (red dashed line).

However, unlike the \textit{sine} function in Eq. \eqref{eq:Sin}, experimental dose-response curves of $\text{CheY}_\text{p}$ are thought to be well approximated by Hill functions rather than Eq. \eqref{eq:Sin} \cite{KeyEndSko06}. There are several possible reasons for this discrepancy. For instance, in our model receptors are either fully sensitive or fully insensitive, and the solution given by Eq. \eqref{eq:Sin} is only valid in the sensitive region and constant otherwise. In reality, receptors have a smooth sensitivity curve spanning the ligand-dissociation constant of the \textit{off} and \textit{on} states (such as $\text{d}F/\text{d} \log(x)$ in \cite{EndWin06}, where $F$ is the receptor free-energy difference between \textit{on} and \textit{off} states). This may lead to a smooth Hill-function-like response.  One way of imposing smooth input-output curves is to introduce the additional constraint of zero first derivatives at the boundary. In this case, however, we only obtain sigmoidal input-output curves without internal switching noise ($\alpha_3$ = 0) (see \textit{S1 Text}, Sec. 1.4.4 and Fig. S3E).
Moreover, very asymmetric (or even bimodal) input distributions might be uncommon in natural environments \cite{Micali14}, potentially favoring log-normal input distributions and hence Hill-function-like responses \cite{Frank13}. Finally, \textit{E. coli} needs to account for many other constraints and the pathway performs other tasks at the same time, such as sensing temperature and pH \cite{OleksiukWinSou11,Sourjik12pH,Tu14PLoSCB}. Hence, the \textit{E. coli} sensory system might be in a suboptimal configuration for transmitting information about chemicals in order to account for all the other tasks.   
In the \textit{S1 Text}, Sec. 1.4.5 we also solve the inverse problem and derive the optimal noise, which leads to an exact Hill function (see Fig. S4). In this case, the predicted input and output noises are not independent anymore.
In summary, the mismatch between the experimental Hill functions and the solution in Eq. \eqref{eq:Sin} is not an artifact of our assumptions on the sources of noise; it emerges when considering independent input and output noise, and when maximizing the mutual information with respect to both the input distribution and the input-output curve. 

%{\color{blue}$x_{\text{on}}= 0.115/c^*$, $x_{\text{off}}= 0.323/c^*$, with $c^*=1 \mu M$ to have a adimentional input} 

\subsection*{Information transmission along the \textit{E. coli} chemotaxis pathway}

Now that we understand the optimization of the mutual information better, we can tackle the second problem: Does the chemotaxis pathway maximise information transmission?
Previous work suggests that the higher the information transmission at the receptors, the higher the drift velocity in the direction of the gradient \cite{Micali14}. Is this finding compatible with the recent observation of the ultrasensitive response of the motor to changes in internal $\text{CheY}_\text{p}$ \cite{YuanBerg13}, or does such a steep response prevent the cell from high information transmission? To answer this question we extend our analysis to the whole chemotaxis pathway. 

\begin{figure}
\includegraphics[width=0.65\columnwidth]{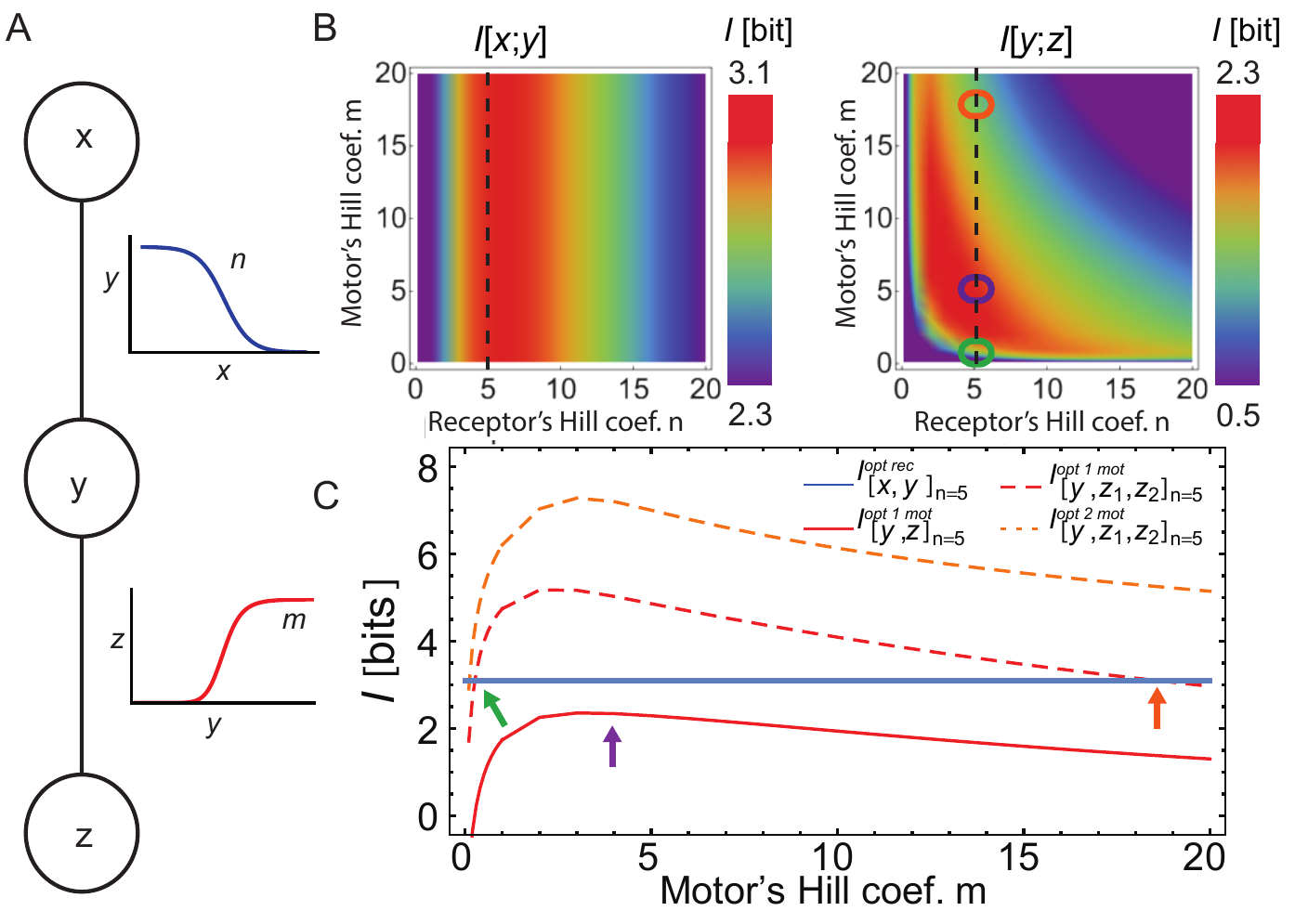}
 \caption{\textbf{Optimal information transmission in the \textit{E. coli} chemotaxis pathway}. (A) Minimal model of the \textit{E. coli} chemotaxis pathway, using two concatenated channels. The extracellular concentration of stimulus $x$ is the input of the receptor channel and $\text{CheY}_\text{p}$, $y$, concentration is the output. For the motor channel, $\text{CheY}_\text{p}$ is the input and the tumble bias $z$ the output. Hill functions with adjustable Hill coefficients $n$ and $m$ represent the input-output curves of the receptor and motor channels, respectively. The dissociation constants for the receptor and motor channels are $k_d^n = 0.189/(1 \mu$M) and $k_d^m = 0.39$ [$Y_T$], respectively.  
 (B) Heat maps of the separately calculated mutual information of the receptor, $\mathcal{I}[x,y]$, (left) and of the motor, $I[y,z]$, (right) channels for a single motor as a function of Hill coefficients $n$ and $m$. Vertical dashed lines correspond to maximal value of $\mathcal{I}[x,y]$ (circles corresponds to arrows in C). 
 (C) The mutual information as function of Hill coefficient $m$ at $n=5$ (dashed black lines in panel B). While for a single motor the optimal mutual information of the receptor channel (\textit{opt rec}, blue solid line) is higher than the optimal mutual information at the motor (\textit{opt mot}, red solid line), increasing the number of motors (see legend) enhances the optimal mutual information of the motor channel (\textit{opt 1 mot}, dashed red line, for two motors). Arrows point to the predicted $m$ values. The purple arrow indicates the optimal $m$ value for a single motor (corresponding to purple circle in B), the green and orange arrows point to the optimal $m$ values for two motors (corresponding to green and orange circles in B). The mutual information is further increased when the optimal mutual information for two motors is calculated (\textit{opt 2 mot}, orange dashed line, see \textit{S1 Text}, Sec 4).} Parameters: $\alpha_{1-3}=10^{-4}$, $\alpha_4=10^{-5}$, $\beta_2=7 \cdot 10^{-4}$, $\beta_3= 7 \cdot 10^{-4}$, and $\beta_4=1 \cdot 10^{-4}$. \label{Fig5}
\end{figure}

We consider a minimal model of two channels: a receptor channel for sensing by the chemoreceptors and a motor channel for the flagellar motors. For the receptor channel, the external chemical concentration $x$ is the input and the internal $\text{CheY}_\text{p}$ concentration, $y$, is the output. For the motor channel, $y$ is the input and the motor clockwise (CW) bias $z$ (for tumble) is the output (Fig. \ref{Fig5}A).  
%{\color{blue}Note that $c^*=1 \mu$M and $Y^*_p = \max \{ \text{CheY} \} = 7.9 \mu M$ are constant to make the input and output dimensionless. We will forget about these constants and refer to the adimentional concentrations $c$ and $Y_p$ for the external stimuli and $\text{CheY}_\text{p}$, respectively.}
To simplify the problem and to closely resemble the experimental dose-response curves, we now restrict the curves to Hill functions, with Hill coefficients $n$ and $m$ for receptors  and motors, respectively.  
The noise expressions for the receptor and motor channels are given by $\sigma_{\text{yT}}=\sqrt{\alpha_1 x G_y^2+ \alpha_2 \bar{y} (1-\bar{y}) + \alpha_3 \bar{y} + \alpha_4}$ and $\sigma_{\text{zT}}=\sqrt{\sigma_{\text{yT}}^2 G_z^2+ \beta_2 \bar{z} (1-\bar{z}) + \beta_3 \bar{z} + \beta_4}$, respectively, where $G_y$ and $G_z$ are the gains of the receptor and motor channels. Parameters $\beta_{2-4}$ represent the noise of the motors and are kept generic due to lack of characterization, but may reflect analogous biological processes including adaptation of the motors \cite{TuBerg12, YuanBerg13, Yuan18BiopJ} (see \textit{S1 Text}, Sec. 3 for further details and robustness of results to changes in $\beta$ values). Due to the immense gain at the motors \cite{YuanBerg13}, we generally have higher noise at the motor then at the receptor (see \textit{S1 Text} Sec. 4.5 for the discussion of the two limits).  
%{\color{blue}RE: is this correct? Can we compare so directly? - GM: I think so, I think it is what we show in Fig. S5. But I am happy to receive suggestions on how to smooth this down.}
Here only $n$ and $m$ are considered adjustable parameters in our model.

The data processing inequality, which characterizes the flow of information in a Markov chain,  
states that, at any additional processing step, information can only be lost, never gained \cite{CoverThomas}. For instantaneous information transmission this means that the mutual information between the external concentration and the motor bias cannot be higher than the minimum of the mutual informations of the receptor and the motor channels, i.e. $\mathcal{I}[x,z]\le \min \left\{ \mathcal{I}[x,y], \mathcal{I}[y,z] \right\}$. A strategy to possibly increase the mutual information is then to maximise the limiting mutual information. 

We start by considering a single motor, represented by a single output $z$. We calculate the maximal mutual information at the receptor (Fig. \ref{Fig5}B, left) and motor (Fig. \ref{Fig5}B, right) channels, dealing with the optimization of the two channels separately. The mutual information at the motor, $\mathcal{I}[y,z]$, always limits the whole information transmission for any Hill coefficient $n$ of the receptors and $m$ of the motor. Hence, single-motor cells should optimize the motor rather than the receptor channel. 
This result is not unexpected for the chemotaxis pathways since the ultrasensitive motor enhances the downstream noise, which is generally larger than the upstream noise (here the total $\text{CheY}_\text{p}$ noise is the input noise for the motor channel). 
The resulting optimal information transmission corresponds to relatively low $n$ and $m$ ($\approx 6$; red area in right panel of Fig. \ref{Fig5}B). 
Experimentally, the Hill coefficient of the receptor channel agrees with our prediction, ranging from $6-12$ in Tar-only cells \cite{EndSouWin2008}. 
In contrast, the ultrasteep motor response curve with $m \approx 20$ is in stark contradiction to our single-motor dose-response model. 
Hence, the single-motor model indicates higher information transmission at the receptors (see Figs. \ref{Fig5}, S6, S7, and \textit{S1 Text}, Secs. 3 and 4.2 for dependence on noise).  However, \textit{E. coli} has multiple motors, which might effect information transmission.

We now extend our model to multiple ($K$) motors, allowing the cell to make multiple measurements of the internal $\text{CheY}_\text{p}$ concentration. 
There are now a single input, $y$, and multiple outputs $z_1,...,z_K$, of the CW biases. The chain rule for the mutual information allows us to calculate $\mathcal{I}[y; z_1,..z_K]$. 
In particular, for the extreme case of fully coupled motors the mutual information does not increase with the growing motor number, $\mathcal{I}[y; z_1,..z_K]= \mathcal{I}[y; z_1]$. In contrast, for completely uncoupled motors (i.e. for motors which are simultaneously independent and conditionally independent given $y$) the mutual information increases with the number of motors, $\mathcal{I}[y; z_1,..z_K]= K \mathcal{I}[y; z_1]$ (see \textit{S1 Text}, Sec. 4.5).
Real motors show evidence of partial coupling \cite{Chemla14Elife}, and thus we assume conditional independent motors, i.e. $p(z_1,..z_K|y) = \Pi^{K}_i p(z_i|y)$. This means that all motors depend on the common $y$ level but can independently select their CW bias. For two motors, the mutual information becomes
$\mathcal{I}[y; z_1, z_2] =  \mathcal{I}[z_1;y] + \mathcal{I}[z_2;y |z_1]  = \mathcal{I}[z_1;y] - H(z_2|y) + H(z_2|z_1)$.
For $K$ motors this is generalized to  
\begin{align}
\mathcal{I}[y; z_1, ... , z_K] = \mathcal{I}[z_1;y] - K H(z_1|y) + \sum_{i=2}^K H(z_i|z_{i-1},...,z_1). 
\label{eq:kmota}
\end{align} 
Using the small noise Gaussian approximation for $p(z_i | y)$, the conditional entropy is given by $H(z_1|y) = \int \text{d} y \ p(y) $ $\log\left( \sqrt{2 \pi \text{e}} \sigma_T(y) \right)$, and $H(z_2|z_1) \approx 0$ (Fig. 5C, see also \textit{S1 Text}, Sec. 4.5). 
Thus, Eq. \eqref{eq:kmota} becomes 
\begin{align}
\mathcal{I}[y; z_1, ... , z_K] = \mathcal{I}[z_1;y] - K \int \text{d} y \ p(y)  \log\left( \sqrt{2 \pi \text{e}} \sigma_T(y) \right).
\end{align}  
We numerically tested that the conditional independence of the motors holds for two motors, despite the fact that motors compete for the binding of internal CheY$_p$ molecules, which can introduce negative correlations (see \textit{S1 Text}, Sec. 4.6 and Fig. S10). 
Therefore, $\mathcal{I}[y; z_1, z_2] > \mathcal{I}[y; z_1]$, and more generally $\mathcal{I}[y; z_1,..z_K]>\mathcal{I}[y; z_1,..z_{K-1}]$ for $K$ conditionally independent motors \cite{Levchenko_11Science}. 

Hence, for conditionally independent motors, the mutual information at the motors will eventually overtake the mutual information at the receptors when the number of motors increases. Consequently, the mutual information at the receptors becomes the limiting factor for information transmission (Fig. \ref{Fig5}C, see \cite{Levchenko_11Science} for the case of Gaussian input distributions).
In other words, for a small number of motors %(or strong coupling) 
the cell has high information transmission at the receptors, which will be wasted at the motors (cf. red and blue solid lines). In contrast, for a large number of motors 
%(or low coupling) 
the information transmission at the motors exceeds the information transmission at the receptors without overall improvement. However, in the intermediate case both receptors and motors equally limit the transmission of information (cf. red dashed and blue solid lines in Fig. \ref{Fig5}C).  
% Note a small number (2-3) of effective motors is suggested in \cite{Chemla14Elife}. 
\textit{E. coli} chemotaxis seems to avoid bottlenecks and to optimally allocate resources the latter case is the most advantageous \cite{tenWolde14PNAS}. Hence, the ultrasteep Hill function of the motor ($m\approx20$) can be explained by this matching of the information transmission at the receptors and motors (orange arrow in Fig. \ref{Fig5}C). 
Note that in addition to the high Hill coefficient $m$ of the motors there is also a corresponding low $m$ solution (green arrow in Fig. \ref{Fig5}C). However, the latter is not robust to changes in $m$, i.e. a small change in $m$ can lead to a drastic reduction of information transmission, which can emerge from varying the number of FliM molecules of the motor \cite{YuanBerg12Nat}. 
In addition, note that the mutual information shown in Fig. \ref{Fig5}C with a red dashed line is calculated assuming that the two motors are optimized separately. However, the mutual information is further increased by maximizing the two motors simultaneously (dashed orange line in Fig. \ref{Fig5}C). Our overall result that a high mutual information can be achieved with a high Hill coefficient of the motors remains valid (see \textit{S1 Text}, Sec. 4). 
In between $m\approx1$ and $\approx20$, the information transmission of the motor is wasted as receptors are information-flow limiting. 
In conclusion, multiple ultrasensitive motors are only useful when motors are sufficiently independent. Any residual coupling among motors may be the result of close motor proximity or mechanical coupling of the flagella.

\section*{Discussion}

This study presents a new approach to maximise the mutual information, particularly suitable for evolving biological systems subject to random mutations and selection. 
Previously, the channel capacity, i.e. the mutual information maximized with respect to the input distribution was widely used for electronic and biological communication channels \cite{TkaCal08a, TkaCal08b,BrunelNadal,KomorowskiStumpf}. However, this method fails to capture possible changes of the internal input-output curve (e.g. by mutations). Furthermore, the mutual information maximized with respect to the input-output curve neglects the biological relevant feedback of the output on the input \cite{DetRamShr00, Micali14}. Here, we reconciled these two approaches by maximizing the mutual information with respect to both the input distribution and the input-output curve for Gaussian channels with small noise. Only when  the total noise is uniform, or when the input or the output noise is negligible, the two approaches are identical. Unlike previous joint optimizations \cite{Tkavcik_09PRE, Walczak_10PRE, Tkavcik_12PRE}, our input-output curves are not restricted to Hill or Hill-like functions. Our adaptive algorithm demonstrates how evolution might implement this iteratively. 

Our analytical solution of the joint optimization provides a number of new insights into optimal information transmission. First, the optimal input distribution is universal, depending only on the input noise. For Berg-and-Purcell type input noise, we specifically obtain $p(x) \sim \sqrt{x}^{-1}$. 
Hence, organisms are optimized for environments in which low intensity stimuli occur with high frequency. This is sensible as their high frequency would compensate for their large relative noise levels. Second, the optimal input-output curve is invariant to up or down scaling of the input noise (parameter $\alpha_1$), which sets the units of the input. 
Hence, only the shape of the input noise (i.e. its functional dependence on input $x$) affects the input-output curve. Third, our optimal input-output curve is rather linear (Figs. 3D and 4A). While this does not match the sigmoidal Hill functions as suggested by models of {\it E. coli} chemotaxis \cite{KeyEndSko06,SouBerg02}, a near linear input-output curve makes best use of a given dynamic range. 
Furthermore, enforcing either zero slope of the input-output curve at the boundaries of the sensitive region or Hill functions as input-output curves leads to assumptions on the noise which are hard to justify biologically. 
Hence, Hill functions are incompatible with independent input and output noise (see Supplementary Information Sec. 1.7 for details).
How can cells actively influence and optimize their distribution of sensory input? Genetic changes in the downstream pathway and motor can clearly change chemotactic behavior and hence the experienced input stimuli. For instance, increases in the motor speed lead to larger changes in stimulus and hence broader distributions of inputs. Similarly, faster adaptation leads to narrower distributions. 
However, the notion that cells influence their microenvironments is most supported by the important role of \textit{niches} in stem cell differentiation, cancer development, gut microbiota, and host-pathogen interactions \cite{Hawkins2016, Tanaka2016, cao2014, plaks2015, messer2017}. Once inside the gut, \textit{E. coli} related pathogen \textit{C. rodentium} in mice (and similarly EPEC/EHEC in humans) injects effector proteins into the epithelial host cells. In response, these cells secrete increased levels of oxygen, allowing in return the pathogen to perform aerobic metabolism \cite{berger2017}.  
Hence, its aerotaxis ability, inherited from \textit{E. coli} based on Aer and Tsr receptors, experiences an increased frequency of oxygen stimuli, which the pathogen actively stimulated. If we take the assumption of maximal information transmission seriously, then cells do not only actively influence but also optimize their environment. %

To apply our information-theoretical approach, we analytically showed that the entire \textit{E. coli} chemotaxis pathway can maximise the instantaneous mutual information between chemical concentration and motor bias despite the ultrasteep dose-response curve of the motors. 
Briefly, ultrasensitive motors do not restrict information transmission, since a collection of motors %(which may provide chemotactic advantages in the soil or animal intestines \cite{Stocker11}) 
boosts information transmission, in addition to providing other chemotactic advantages in the soil or animal intestine \cite{Stocker11}.
In particular, our model identifies the number of motors and their conditional independence as key quantities to transmit large amounts of information in peritrichous bacteria.  %despite the high sensitivity. 

What is the additional information at the motors used for if the ultimate behavioral output is just binary runs and tumbles? We speculate that the tumble angle, torque, and filament handedness could be regulated \cite{Turner00,Berg07torque}. Indeed, real-time imaging of \textit{E.coli} with fluorescent flagella showed that the tumble angles increased with the number of clockwise-turning motors, allowing for differential cell responses \cite{Turner00}. 
Having non-identical motors with different Hill coefficients and thresholds may further increase the information transmission (e.g. as produced by different number of FliM in the motor ring) \cite{Tkavcik_09PRE} but this may not be feasible in the bacterial chemotaxis pathway, as the adapted activity set the operating point of the motors. For instance, different threshold values for the motor would lead to some motor always rotating clockwise and counter-clockwise. 
Our model also makes the prediction that chemotactic bacteria with a single motor should prefer a relatively low Hill coefficient at the motor or multiple response regulators feeding into a motor with a high Hill coefficient to highly transmit information. This prediction could be tested with the uni-flagellated bacterial species, such as \textit{Rhodobacter sphaeroides}, \textit{Pseudomonas aerugiuosa} or monotrichous marine bacteria \cite{SourjikArmitageRev,Stocker11}. In support of our theory, \textit{R. sphaeroides} is known to have multiple CheY's \cite{PortWadArm2008, Porter11}.

While applicable to many biological systems, our model makes a number of simplifications (in addition to assumptions on noise and receptor sensitivity). 
Our results are based on the independent maximizations of the receptor and motor channels. However in \textit{SI text}, Sec. 4.5, we discuss the general case, providing estimes of the mutual information $\mathcal{I}[x;z]$, between ligand input and final motor output. Our analysis suggests, once again, that high Hill coefficient for multiple motors can support high information transmission. In particular, we analytically identify two expected limits, (i) when the receptor noise is much smaller than the motor noise, we obtain $\mathcal{I}[x;z] \approx \mathcal{I}[y;z]$, and (ii) when the motor noise is smaller than the receptor noise, we have $\mathcal{I}[x;z] \approx \mathcal{I}[x;y]$. 
Our analysis over the chemotactic pathway primarily focuses on the Hill coefficient. However, the dissociation constant $k^m_d$ of the motor response is known to be larger than the adapted CheY$_\text{p}$ concentration. In \textit{SI text}, Sec. 4.5.4, we explicitly study the role of the dissociation constant $k^m_d$, and found a relative weak dependence of the mutual information on it. We also found that after fixing the Hill coefficient $m$ and using the optimized output distribution of the receptor channel, the $k^m_d$ that maximizes the mutual information at the motor matches the experimentally measured value (which is larger than the adapted CheY$_\text{p}$ level, see Fig. S8).   

Another simplification is that our model deals with instantaneous information transmission, and hence does not explicitly include any history dependence \cite{TuShimBerg08, TosttenWolde09, ShimTuBerg10,tenWolde15prl}. 
Hence, our approach should be highly suitable for the slow genetic response in quorum sensing \cite{MehGoyWin2009,taillefumier2015}. 
In this system, the input-output relation has been measured but input distributions were simply guessed, and not predicted. Another area of application is eukaryotic chemotaxis, where cells move slowly while actively shaping their chemical gradient by ligand secretion \cite{DePalo17} and degradation \cite{Tweedy16}. In all these examples, the input distributions and cell behaviors need to match the input-output relations to allow for optimal information gathering.
Nevertheless, our model is valid for information transmission by initial transient chemotactic responses, and as this applies anywhere in the gradient, our model describes chemotaxis even including adaptation \cite{Micali14}. 
We expect that our model even works in relatively steep gradients, where, in addition to adaptation, long-history effects are important, such as caused by receptor saturation and rotational diffusion \cite{Micali17}.  
The main assumption in \cite{Micali14} is that gradients can be linearized over the range of input distributions. However, we do not assume small Gaussian-distributed inputs. A drawback of our model is that we neglect any cell-to-cell variability, which can be substantial \cite{ColinSourjik17, Keegstra2017}, so that in effect our theory focuses on a certain subpopulation of cells. This cell-to-cell variability may lead to advantages in terms of bet-hedging strategies not directly related to information processing \cite{EmonetFrankel_eLife15}.   

In conclusion, we provided a biologically inspired adaptive algorithm with analytical solution for complex problems in information transmission in sensory systems. Future studies may need to account for time-dependencies explicitly, including adaptation both at the receptors and motors. This may be achieved by considering trajectories of molecular concentrations and/or cell behavior, which may also help establish a link between chemotactic performance (e.g. drift velocity), information transmission, and energetic cost of chemotaxis. This link may show interesting tradeoffs and new design principles \cite{Micali16}. A methodological contribution might be necessary as calculation of the mutual information based on trajectories are hampered by the high-dimensional phase space of all possible trajectories.

%\section*{Conclusion}

\section*{Materials and methods}
\subsection*{Maximization of mutual information with respect to input distribution and input-output curve}
To maximize Eq. \eqref{eq:MI2} assuming the noise in Eq. \eqref{eq:noise}, we focus on the integrand $\mathcal{L}(x,p,\bar{y},\bar{y}')= p \, \text{log}_2 \left[ \frac{\sqrt{2 \pi e} \sigma_\text{T} }{\bar{y}'} p \right]$ and use the well-known Lagrange formalism from calculus of variations, where $x$ is the independent variable while $p=p(x)$, $\bar{y}=\bar{y}(x)$ and $\bar{y}'=\bar{y}'(x)$ are the dependent variables. Note that $p'(x)$ is not appearing in the Lagrangian $\mathcal{L}$. To find the maximum of $I$ with respect to $p$ and $\bar{y}$, we need to solve Eqs. \eqref{eq:sist1} and \eqref{eq:sist2} together. 

Eq. \eqref{eq:solSy1} is the solution of Eq. \eqref{eq:sist1}, which can be rewritten through differentiation as  
\begin{align}
\label{eq:diffSist1}
\frac{p'}{p}= \frac{\bar{y}''}{\bar{y}'} - \frac{\sigma_\text{T}'}{\sigma_\text{T}}.  
\end{align}
To derive Eq. \eqref{eq:solSy2}, we evaluate the following derivatives 
\begin{align}
%\label{eq:dldy}
\frac{\partial \mathcal{L}}{\partial \bar{y}} &=  \frac{p}{\sigma_\text{T}} \frac{\partial \sigma_\text{T}}{\partial \bar{y}}, \\
\frac{\partial \mathcal{L}}{\partial \bar{y}'} &= \frac{p}{\sigma_\text{T}} \frac{\partial \sigma_\text{T}}{\partial \bar{y}'} -\frac{p}{\bar{y}'}, \\
\label{eq:ddxdldyp}
\frac{d}{dx}\frac{\partial \mathcal{L}}{\partial \bar{y}'} &= -\frac{p'}{\bar{y'}}+\frac{p \bar{y}''}{\bar{y}'^2} -\frac{p'}{\sigma_\text{T}}\frac{\partial \sigma_\text{T}}{\partial \bar{y}'} +\frac{p \sigma_\text{T}'}{\sigma_\text{T}^2}\frac{\partial \sigma_\text{T}}{\partial \bar{y}} 
+ \frac{p}{\sigma_\text{T}}\frac{d}{dx} \frac{\partial \sigma_\text{T}}{\partial \bar{y}'}.
\end{align}
%By equating \eqref{eq:dldy} to \eqref{eq:ddxdldyp}, and using Eq. \eqref{eq:diffSist1}, we get
Using Eqs. \eqref{eq:diffSist1}-\eqref{eq:ddxdldyp}, Eq. \eqref{eq:sist2} becomes
\begin{align}
\label{eq:SolGenN}
2 \frac{\bar{y}''}{\bar{y}'} - 2 \frac{\sigma_\text{T}'}{\sigma_\text{T}} + \frac{\frac{d}{dx}  \frac{\partial \sigma_\text{T}}{\partial \bar{y}'}}{\frac{\partial \sigma_\text{T}}{\partial \bar{y}'}} = -\frac{1}{\bar{y}'} \frac{\frac{\partial \sigma_\text{T}}{\partial x}}{\frac{\partial \sigma_\text{T}}{\partial \bar{y}'}}.
\end{align}
%Integrating, 
%\begin{align}
%\log \frac{\bar{y}'^2 \frac{\partial \sigma_\text{T}}{\partial \bar{y}'}}{\sigma_\text{T}}=-
%\end{align}
Now assuming independent cell-external and internal noises as in Eq. \eqref{eq:noise}, Eq. \eqref{eq:SolGenN} becomes Eq. \eqref{eq:solSy2}.

\section*{Author contributions statement}
Both authors (GM and RGE) designed the study. GM performed the research. Both analyzed results and data, and wrote the paper.

\section*{Competing interests statement}
GM and RGE declare that there is no conflict of interest, neither financial nor non-financial.

\section*{Acknowledgments}
We thank Chiu Fan Lee, Thomas Ouldridge, Michael Stumpf and Peter Swain for valuable comments on the manuscript, Judith Armitage for valuable discussions about $\textit{Rhodobacter sphaeroides}$, and Victor Sourjik for collaborating with us on related projects. This work was supported by European Research Council Starting-Grant N. 280492-PPHPI. Additional support came from the Biotechnological and Biological Sciences Research Council grant BB/G000131/1 (RGE) and the Swiss National Science Foundation grant nr. 31003A\_169978 to Martin Ackermann (GM). 

\nolinenumbers

% Either type in your references using
% \begin{thebibliography}{}
% Text
% \end{thebibliography}
%
% or
%
% Compile your BiBTeX database using our plos2015.bst
% style file and paste the contents of your .bbl file
% here. See http://journals.plos.org/plosone/s/latex for 
\bibliographystyle{naturemag}
\bibliography{all}

\begin{thebibliography}{10}
\expandafter\ifx\csname url\endcsname\relax
  \def\url#1{\texttt{#1}}\fi
\expandafter\ifx\csname urlprefix\endcsname\relax\def\urlprefix{URL }\fi
\providecommand{\bibinfo}[2]{#2}
\providecommand{\eprint}[2][]{\url{#2}}

\bibitem{Swain_msbREW}
\bibinfo{author}{Perkins, T.~J.} \& \bibinfo{author}{Swain, P.~S.}
\newblock \bibinfo{title}{Strategies for cellular decision-making}.
\newblock \emph{\bibinfo{journal}{Mol Syst Biol}} \textbf{\bibinfo{volume}{5}},
  \bibinfo{pages}{326--326} (\bibinfo{year}{2009}).

\bibitem{Vergassola2007infotaxis}
\bibinfo{author}{Vergassola, M.}, \bibinfo{author}{Villermaux, E.} \&
  \bibinfo{author}{Shraiman, B.~I.}
\newblock \bibinfo{title}{‘infotaxis’ as a strategy for searching without
  gradients}.
\newblock \emph{\bibinfo{journal}{Nature}} \textbf{\bibinfo{volume}{445}},
  \bibinfo{pages}{406} (\bibinfo{year}{2007}).

\bibitem{Lander13}
\bibinfo{author}{Lander, A.~D.}
\newblock \bibinfo{title}{How cells know where they are}.
\newblock \emph{\bibinfo{journal}{Science}} \textbf{\bibinfo{volume}{339}},
  \bibinfo{pages}{923--927} (\bibinfo{year}{2013}).

\bibitem{Taylor2007information}
\bibinfo{author}{Taylor, S.~F.}, \bibinfo{author}{Tishby, N.} \&
  \bibinfo{author}{Bialek, W.}
\newblock \bibinfo{title}{Information and fitness}.
\newblock \emph{\bibinfo{journal}{arXiv preprint}}
  \textbf{\bibinfo{volume}{0712.4382}} (\bibinfo{year}{2007}).

\bibitem{Rivoire_Leibler_11}
\bibinfo{author}{Rivoire, O.} \& \bibinfo{author}{Leibler, S.}
\newblock \bibinfo{title}{The value of information for populations in varying
  environments}.
\newblock \emph{\bibinfo{journal}{J Stat Phys}} \textbf{\bibinfo{volume}{142}},
  \bibinfo{pages}{1124--1166} (\bibinfo{year}{2011}).

\bibitem{Berg00}
\bibinfo{author}{Berg, H.~C.}
\newblock \bibinfo{title}{Motile behavior of bacteria.}
\newblock \emph{\bibinfo{journal}{Phys Today}} \textbf{\bibinfo{volume}{53}},
  \bibinfo{pages}{24--29} (\bibinfo{year}{2000}).

\bibitem{EndresBook13}
\bibinfo{author}{Endres, R.}
\newblock \emph{\bibinfo{title}{Physical Principles in Sensing and Signaling:
  With an Introduction to Modeling in Biology}} (\bibinfo{publisher}{Oxford
  University Press}, \bibinfo{year}{2013}).

\bibitem{Shannon}
\bibinfo{author}{Shannon, C.~E.}
\newblock \bibinfo{title}{A mathematical theory of communication.}
\newblock \emph{\bibinfo{journal}{Bell Syst Tech J}}
  \textbf{\bibinfo{volume}{27}}, \bibinfo{pages}{379--423}
  (\bibinfo{year}{1948}).

\bibitem{SwainRevInfo14}
\bibinfo{author}{Bowsher, C.~G.} \& \bibinfo{author}{Swain, P.~S.}
\newblock \bibinfo{title}{Environmental sensing, information transfer, and
  cellular decision-making}.
\newblock \emph{\bibinfo{journal}{Curr Opin Biotech}}
  \textbf{\bibinfo{volume}{28}}, \bibinfo{pages}{149--155}
  (\bibinfo{year}{2014}).

\bibitem{Tkavcik_11Rev}
\bibinfo{author}{Tka{\v{c}}ik, G.} \& \bibinfo{author}{Walczak, A.~M.}
\newblock \bibinfo{title}{Information transmission in genetic regulatory
  networks: a review}.
\newblock \emph{\bibinfo{journal}{Journal of Physics: Condensed Matter}}
  \textbf{\bibinfo{volume}{23}}, \bibinfo{pages}{153102}
  (\bibinfo{year}{2011}).

\bibitem{Tkacik_16Rev}
\bibinfo{author}{Tka{\v{c}}ik, G.} \& \bibinfo{author}{Bialek, W.}
\newblock \bibinfo{title}{Information processing in living systems}.
\newblock \emph{\bibinfo{journal}{Annual Review of Condensed Matter Physics}}
  \textbf{\bibinfo{volume}{7}}, \bibinfo{pages}{89--117}
  (\bibinfo{year}{2016}).

\bibitem{TkaCal08a}
\bibinfo{author}{Tka{\v{c}}ik, G.}, \bibinfo{author}{{Callan Jr}, C.~G.} \&
  \bibinfo{author}{Bialek, W.}
\newblock \bibinfo{title}{{Information flow and optimization in transcriptional
  regulation}}.
\newblock \emph{\bibinfo{journal}{Proc Natl Acad Sci U S A}}
  \textbf{\bibinfo{volume}{105}}, \bibinfo{pages}{12265--12270}
  (\bibinfo{year}{2008}).

\bibitem{TkaCal08b}
\bibinfo{author}{Tka{\v{c}}ik, G.}, \bibinfo{author}{{Callan Jr}, C.~G.} \&
  \bibinfo{author}{Bialek, W.}
\newblock \bibinfo{title}{Information capacity of genetic regulatory elements}.
\newblock \emph{\bibinfo{journal}{Phys Rev E}} \textbf{\bibinfo{volume}{78}},
  \bibinfo{pages}{011910} (\bibinfo{year}{2008}).

\bibitem{DetRamShr00}
\bibinfo{author}{Detwiler, P.~B.}, \bibinfo{author}{Ramanathan, S.},
  \bibinfo{author}{Sengupta, A.} \& \bibinfo{author}{Shraiman, B.~I.}
\newblock \bibinfo{title}{Engineering aspects of enzymatic signal transduction:
  photoreceptors in the retina.}
\newblock \emph{\bibinfo{journal}{Biophys J}} \textbf{\bibinfo{volume}{79}},
  \bibinfo{pages}{2801--2817} (\bibinfo{year}{2000}).

\bibitem{Micali14}
\bibinfo{author}{Clausznitzer, D.}, \bibinfo{author}{Micali, G.},
  \bibinfo{author}{Neumann, S.}, \bibinfo{author}{Sourjik, V.} \&
  \bibinfo{author}{Endres, R.~G.}
\newblock \bibinfo{title}{Predicting chemical environments of bacteria from
  receptor signaling}.
\newblock \emph{\bibinfo{journal}{PLoS Comput Biol}}
  \textbf{\bibinfo{volume}{10}}, \bibinfo{pages}{e1003870}
  (\bibinfo{year}{2014}).

\bibitem{Tkavcik_09PRE}
\bibinfo{author}{Tka{\v{c}}ik, G.}, \bibinfo{author}{Walczak, A.~M.} \&
  \bibinfo{author}{Bialek, W.}
\newblock \bibinfo{title}{Optimizing information flow in small genetic
  networks}.
\newblock \emph{\bibinfo{journal}{Physical Review E}}
  \textbf{\bibinfo{volume}{80}}, \bibinfo{pages}{031920}
  (\bibinfo{year}{2009}).

\bibitem{Walczak_10PRE}
\bibinfo{author}{Walczak, A.~M.}, \bibinfo{author}{Tka{\v{c}}ik, G.} \&
  \bibinfo{author}{Bialek, W.}
\newblock \bibinfo{title}{Optimizing information flow in small genetic
  networks. ii. feed-forward interactions}.
\newblock \emph{\bibinfo{journal}{Physical Review E}}
  \textbf{\bibinfo{volume}{81}}, \bibinfo{pages}{041905}
  (\bibinfo{year}{2010}).

\bibitem{Tkavcik_12PRE}
\bibinfo{author}{Tka{\v{c}}ik, G.}, \bibinfo{author}{Walczak, A.~M.} \&
  \bibinfo{author}{Bialek, W.}
\newblock \bibinfo{title}{Optimizing information flow in small genetic
  networks. iii. a self-interacting gene}.
\newblock \emph{\bibinfo{journal}{Physical Review E}}
  \textbf{\bibinfo{volume}{85}}, \bibinfo{pages}{041903}
  (\bibinfo{year}{2012}).

\bibitem{bernardo1979reference}
\bibinfo{author}{Bernardo, J.~M.}
\newblock \bibinfo{title}{Reference posterior distributions for {Bayesian}
  inference}.
\newblock \emph{\bibinfo{journal}{J R Stat Soc Ser B Stat Methodol}}
  \textbf{\bibinfo{volume}{41}}, \bibinfo{pages}{113--147}
  (\bibinfo{year}{1979}).

\bibitem{BrunelNadal}
\bibinfo{author}{Brunel, N.} \& \bibinfo{author}{Nadal, J.~P.}
\newblock \bibinfo{title}{Mutual information, {Fisher} information, and
  population coding}.
\newblock \emph{\bibinfo{journal}{Neural Comput}}
  \textbf{\bibinfo{volume}{10}}, \bibinfo{pages}{1731--1757}
  (\bibinfo{year}{1998}).

\bibitem{YuanBerg13}
\bibinfo{author}{Yuan, J.} \& \bibinfo{author}{Berg, H.~C.}
\newblock \bibinfo{title}{Ultrasensitivity of an adaptive bacterial motor}.
\newblock \emph{\bibinfo{journal}{J Mol Biol}} \textbf{\bibinfo{volume}{425}},
  \bibinfo{pages}{1760--1764} (\bibinfo{year}{2013}).

\bibitem{Dufour14}
\bibinfo{author}{Dufour, Y.~S.}, \bibinfo{author}{Fu, X.},
  \bibinfo{author}{Hernandez-Nunez, L.} \& \bibinfo{author}{Emonet, T.}
\newblock \bibinfo{title}{Limits of feedback control in bacterial chemotaxis}.
\newblock \emph{\bibinfo{journal}{PLoS Comput Biol}}
  \textbf{\bibinfo{volume}{10}}, \bibinfo{pages}{e1003694}
  (\bibinfo{year}{2014}).

\bibitem{Vergassola16}
\bibinfo{author}{Wong-Ng, J.}, \bibinfo{author}{Melbinger, A.},
  \bibinfo{author}{Celani, A.} \& \bibinfo{author}{Vergassola, M.}
\newblock \bibinfo{title}{The role of adaptation in bacterial speed races}.
\newblock \emph{\bibinfo{journal}{PLoS Comput Biol}}
  \textbf{\bibinfo{volume}{12}}, \bibinfo{pages}{e1004974}
  (\bibinfo{year}{2016}).

\bibitem{Micali17}
\bibinfo{author}{Micali, G.}, \bibinfo{author}{Colin, R.},
  \bibinfo{author}{Sourjik, V.} \& \bibinfo{author}{Endres, R.~G.}
\newblock \bibinfo{title}{Drift and behavior of e. coli cells}.
\newblock \emph{\bibinfo{journal}{Biophysical Journal}}
  \textbf{\bibinfo{volume}{113}}, \bibinfo{pages}{2321 -- 2325}
  (\bibinfo{year}{2017}).

\bibitem{Skoge_Meir_Win_PRL11}
\bibinfo{author}{Skoge, M.}, \bibinfo{author}{Meir, Y.} \&
  \bibinfo{author}{Wingreen, N.~S.}
\newblock \bibinfo{title}{Dynamics of cooperativity in chemical sensing among
  cell-surface receptors}.
\newblock \emph{\bibinfo{journal}{Phys Rev Lett}}
  \textbf{\bibinfo{volume}{107}}, \bibinfo{pages}{178101--178101}
  (\bibinfo{year}{2011}).

\bibitem{ThomasP2016}
\bibinfo{author}{Thomas, P.~J.} \& \bibinfo{author}{Eckford, A.~W.}
\newblock \bibinfo{title}{Capacity of a simple intercellular signal
  transduction channel}.
\newblock \emph{\bibinfo{journal}{IEEE Transactions on Information Theory}}
  \textbf{\bibinfo{volume}{62}}, \bibinfo{pages}{7358--7382}
  (\bibinfo{year}{2016}).

\bibitem{SouBerg02a}
\bibinfo{author}{Sourjik, V.} \& \bibinfo{author}{Berg, H.~C.}
\newblock \bibinfo{title}{Receptor sensitivity in bacterial chemotaxis.}
\newblock \emph{\bibinfo{journal}{Proc Natl Acad Sci U S A}}
  \textbf{\bibinfo{volume}{99}}, \bibinfo{pages}{123--127}
  (\bibinfo{year}{2002}).

\bibitem{YuanBerg12Nat}
\bibinfo{author}{Yuan, J.}, \bibinfo{author}{Branch, R.~W.},
  \bibinfo{author}{Hosu, B.~G.} \& \bibinfo{author}{Berg, H.~C.}
\newblock \bibinfo{title}{Adaptation at the output of the chemotaxis signalling
  pathway}.
\newblock \emph{\bibinfo{journal}{Nature}} \textbf{\bibinfo{volume}{484}},
  \bibinfo{pages}{233--236} (\bibinfo{year}{2012}).

\bibitem{Nemenman_14Review}
\bibinfo{author}{Levchenko, A.} \& \bibinfo{author}{Nemenman, I.}
\newblock \bibinfo{title}{Cellular noise and information transmission}.
\newblock \emph{\bibinfo{journal}{Curr Opin Biotechnol}}
  \textbf{\bibinfo{volume}{28}}, \bibinfo{pages}{156--164}
  (\bibinfo{year}{2014}).

\bibitem{McMahon_14}
\bibinfo{author}{Mc~Mahon, S.~S.} \emph{et~al.}
\newblock \bibinfo{title}{Information theory and signal transduction systems:
  From molecular information processing to network inference}.
\newblock \emph{\bibinfo{journal}{Semin Cell Dev Biol}}
  \textbf{\bibinfo{volume}{35C}}, \bibinfo{pages}{98--108}
  (\bibinfo{year}{2014}).

\bibitem{Micali16}
\bibinfo{author}{Micali, G.} \& \bibinfo{author}{Endres, R.~G.}
\newblock \bibinfo{title}{Bacterial chemotaxis: information processing,
  thermodynamics, and behavior}.
\newblock \emph{\bibinfo{journal}{Curr Opin Microbiol}}
  \textbf{\bibinfo{volume}{30}}, \bibinfo{pages}{8--15} (\bibinfo{year}{2016}).

\bibitem{Lau81}
\bibinfo{author}{Laughlin, S.}
\newblock \bibinfo{title}{A simple coding procedure enhances a neuron's
  information capacity}.
\newblock \emph{\bibinfo{journal}{Z Naturforsch C}}
  \textbf{\bibinfo{volume}{36}}, \bibinfo{pages}{910--912}
  (\bibinfo{year}{1981}).

\bibitem{JeffreysP}
\bibinfo{author}{Jeffreys, H.}
\newblock \bibinfo{title}{An invariant form for the prior probability in
  estimation problems}.
\newblock \emph{\bibinfo{journal}{P Roy Soc Lond A Mat}}
  \textbf{\bibinfo{volume}{186}}, \bibinfo{pages}{453--461}
  (\bibinfo{year}{1946}).

\bibitem{Blahut72}
\bibinfo{author}{Blahut, R.}
\newblock \bibinfo{title}{Computation of channel capacity and rate-distortion
  functions}.
\newblock \emph{\bibinfo{journal}{IEEE transactions on Information Theory}}
  \textbf{\bibinfo{volume}{18}}, \bibinfo{pages}{460--473}
  (\bibinfo{year}{1972}).

\bibitem{KomorowskiStumpf}
\bibinfo{author}{Komorowski, M.} \emph{et~al.}
\newblock \bibinfo{title}{Analog nitrogen sensing in escherichia coli enables
  high fidelity information processing}.
\newblock \emph{\bibinfo{journal}{bioRxiv}} \bibinfo{pages}{015792}
  (\bibinfo{year}{2015}).

\bibitem{KeyEndSko06}
\bibinfo{author}{Keymer, J.~E.}, \bibinfo{author}{Endres, R.~G.},
  \bibinfo{author}{Skoge, M.}, \bibinfo{author}{Meir, Y.} \&
  \bibinfo{author}{Wingreen, N.~S.}
\newblock \bibinfo{title}{Chemosensing in escherichia coli: two regimes of
  two-state receptors}.
\newblock \emph{\bibinfo{journal}{Proc Natl Acad Sci U S A}}
  \textbf{\bibinfo{volume}{103}}, \bibinfo{pages}{1786--1791}
  (\bibinfo{year}{2006}).

\bibitem{EndSouWin2008}
\bibinfo{author}{Endres, R.~G.} \emph{et~al.}
\newblock \bibinfo{title}{Variable sizes of \emph{Escherichia coli}
  chemoreceptor signaling teams.}
\newblock \emph{\bibinfo{journal}{Mol Syst Biol}} \textbf{\bibinfo{volume}{4}},
  \bibinfo{pages}{211} (\bibinfo{year}{2008}).

\bibitem{TosttenWolde09}
\bibinfo{author}{Tostevin, F.} \& \bibinfo{author}{ten Wolde, P.~R.}
\newblock \bibinfo{title}{Mutual information between input and output
  trajectories of biochemical networks}.
\newblock \emph{\bibinfo{journal}{Phys Rev Lett}}
  \textbf{\bibinfo{volume}{102}}, \bibinfo{pages}{218101--218101}
  (\bibinfo{year}{2009}).

\bibitem{tenWolde15prl}
\bibinfo{author}{Becker, N.~B.}, \bibinfo{author}{Mugler, A.} \&
  \bibinfo{author}{ten Wolde, P.~R.}
\newblock \bibinfo{title}{Optimal prediction by cellular signaling networks}.
\newblock \emph{\bibinfo{journal}{Phys Rev Lett}}
  \textbf{\bibinfo{volume}{115}}, \bibinfo{pages}{258103}
  (\bibinfo{year}{2015}).

\bibitem{ClauEnd11BMC}
\bibinfo{author}{Clausznitzer, D.} \& \bibinfo{author}{Endres, R.~G.}
\newblock \bibinfo{title}{Noise characteristics of the {Escherichia} coli
  rotary motor}.
\newblock \emph{\bibinfo{journal}{BMC Syst Biol}} \textbf{\bibinfo{volume}{5}},
  \bibinfo{pages}{151--151} (\bibinfo{year}{2011}).

\bibitem{BergPur77}
\bibinfo{author}{Berg, H.~C.} \& \bibinfo{author}{Purcell, E.~M.}
\newblock \bibinfo{title}{Physics of chemoreception}.
\newblock \emph{\bibinfo{journal}{Biophys J}} \textbf{\bibinfo{volume}{20}},
  \bibinfo{pages}{193--219} (\bibinfo{year}{1977}).

\bibitem{Laub15_rev}
\bibinfo{author}{Salazar, M.~E.} \& \bibinfo{author}{Laub, M.~T.}
\newblock \bibinfo{title}{Temporal and evolutionary dynamics of two-component
  signaling pathways}.
\newblock \emph{\bibinfo{journal}{Curr Opin Microbiol}}
  \textbf{\bibinfo{volume}{24}}, \bibinfo{pages}{7--14} (\bibinfo{year}{2015}).

\bibitem{EndWin06}
\bibinfo{author}{Endres, R.~G.} \& \bibinfo{author}{Wingreen, N.~S.}
\newblock \bibinfo{title}{Precise adaptation in bacterial chemotaxis through
  “assistance neighborhoods”}.
\newblock \emph{\bibinfo{journal}{Proc Natl Acad Sci U S A}}
  \textbf{\bibinfo{volume}{103}}, \bibinfo{pages}{13040--13044}
  (\bibinfo{year}{2006}).

\bibitem{Frank13}
\bibinfo{author}{Frank, S.~A.}
\newblock \bibinfo{title}{Input-output relations in biological systems:
  measurement, information and the {Hill} equation}.
\newblock \emph{\bibinfo{journal}{Biol Direct}} \textbf{\bibinfo{volume}{8}},
  \bibinfo{pages}{31--31} (\bibinfo{year}{2013}).

\bibitem{OleksiukWinSou11}
\bibinfo{author}{Oleksiuk, O.} \emph{et~al.}
\newblock \bibinfo{title}{Thermal robustness of signaling in bacterial
  chemotaxis}.
\newblock \emph{\bibinfo{journal}{Cell}} \textbf{\bibinfo{volume}{145}},
  \bibinfo{pages}{312--321} (\bibinfo{year}{2011}).

\bibitem{Sourjik12pH}
\bibinfo{author}{Yang, Y.} \& \bibinfo{author}{Sourjik, V.}
\newblock \bibinfo{title}{Opposite responses by different chemoreceptors set a
  tunable preference point in {Escherichia coli} {pH} taxis}.
\newblock \emph{\bibinfo{journal}{Mol Microbiol}}
  \textbf{\bibinfo{volume}{86}}, \bibinfo{pages}{1482--1489}
  (\bibinfo{year}{2012}).

\bibitem{Tu14PLoSCB}
\bibinfo{author}{Hu, B.} \& \bibinfo{author}{Tu, Y.}
\newblock \bibinfo{title}{Behaviors and strategies of bacterial navigation in
  chemical and nonchemical gradients}.
\newblock \emph{\bibinfo{journal}{PLoS Comput Biol}}
  \textbf{\bibinfo{volume}{10}}, \bibinfo{pages}{e1003672}
  (\bibinfo{year}{2014}).

\bibitem{TuBerg12}
\bibinfo{author}{Tu, Y.} \& \bibinfo{author}{Berg, H.~C.}
\newblock \bibinfo{title}{Tandem adaptation with a common design in escherichia
  coli chemotaxis}.
\newblock \emph{\bibinfo{journal}{Journal of molecular biology}}
  \textbf{\bibinfo{volume}{423}}, \bibinfo{pages}{782--788}
  (\bibinfo{year}{2012}).

\bibitem{Yuan18BiopJ}
\bibinfo{author}{Zhang, C.}, \bibinfo{author}{He, R.}, \bibinfo{author}{Zhang,
  R.} \& \bibinfo{author}{Yuan, J.}
\newblock \bibinfo{title}{Motor adaptive remodeling speeds up bacterial
  chemotactic adaptation}.
\newblock \emph{\bibinfo{journal}{Biophys J}} \textbf{\bibinfo{volume}{114}},
  \bibinfo{pages}{1225--1231} (\bibinfo{year}{2018}).

\bibitem{CoverThomas}
\bibinfo{author}{Cover, T.~M.} \& \bibinfo{author}{Thomas, J.~A.}
\newblock \emph{\bibinfo{title}{{Elements of Information Theory (Wiley Series
  in Telecommunications and Signal Processing)}}}
  (\bibinfo{publisher}{Wiley-Interscience}, \bibinfo{year}{1991}),
  \bibinfo{edition}{99} edn.

\bibitem{Chemla14Elife}
\bibinfo{author}{Mears, P.~J.}, \bibinfo{author}{Koirala, S.},
  \bibinfo{author}{Rao, C.~V.}, \bibinfo{author}{Golding, I.} \&
  \bibinfo{author}{Chemla, Y.~R.}
\newblock \bibinfo{title}{Escherichia coli swimming is robust against
  variations in flagellar number}.
\newblock \emph{\bibinfo{journal}{eLife}} \textbf{\bibinfo{volume}{3}},
  \bibinfo{pages}{e01916} (\bibinfo{year}{2014}).

\bibitem{Levchenko_11Science}
\bibinfo{author}{Cheong, R.}, \bibinfo{author}{Rhee, A.},
  \bibinfo{author}{Wang, C.~J.}, \bibinfo{author}{Nemenman, I.} \&
  \bibinfo{author}{Levchenko, A.}
\newblock \bibinfo{title}{Information transduction capacity of noisy
  biochemical signaling networks}.
\newblock \emph{\bibinfo{journal}{Science}} \textbf{\bibinfo{volume}{334}},
  \bibinfo{pages}{354--358} (\bibinfo{year}{2011}).

\bibitem{tenWolde14PNAS}
\bibinfo{author}{Govern, C.~C.} \& \bibinfo{author}{ten Wolde, P.~R.}
\newblock \bibinfo{title}{Optimal resource allocation in cellular sensing
  systems}.
\newblock \emph{\bibinfo{journal}{Proc Natl Acad Sci U S A}}
  \textbf{\bibinfo{volume}{111}}, \bibinfo{pages}{17486--17491}
  (\bibinfo{year}{2014}).

\bibitem{SouBerg02}
\bibinfo{author}{Sourjik, V.} \& \bibinfo{author}{Berg, H.~C.}
\newblock \bibinfo{title}{{Binding of the \emph{Escherichia coli} response
  regulator {CheY} to its target measured in vivo by fluorescence resonance
  energy transfer}}.
\newblock \emph{\bibinfo{journal}{Proc Natl Acad Sci U S A}}
  \textbf{\bibinfo{volume}{99}}, \bibinfo{pages}{12669--12674}
  (\bibinfo{year}{2002}).

\bibitem{Hawkins2016}
\bibinfo{author}{Hawkins, E.~D.} \emph{et~al.}
\newblock \bibinfo{title}{T-cell acute leukaemia exhibits dynamic interactions
  with bone marrow microenvironments}.
\newblock \emph{\bibinfo{journal}{Nature}} \textbf{\bibinfo{volume}{538}},
  \bibinfo{pages}{518} (\bibinfo{year}{2016}).

\bibitem{Tanaka2016}
\bibinfo{author}{Tanaka, M.} \emph{et~al.}
\newblock \bibinfo{title}{Identification of anti-cancer chemical compounds
  using xenopus embryos}.
\newblock \emph{\bibinfo{journal}{Cancer science}}
  \textbf{\bibinfo{volume}{107}}, \bibinfo{pages}{803--811}
  (\bibinfo{year}{2016}).

\bibitem{cao2014}
\bibinfo{author}{Cao, Z.} \emph{et~al.}
\newblock \bibinfo{title}{Angiocrine factors deployed by tumor vascular niche
  induce b cell lymphoma invasiveness and chemoresistance}.
\newblock \emph{\bibinfo{journal}{Cancer cell}} \textbf{\bibinfo{volume}{25}},
  \bibinfo{pages}{350--365} (\bibinfo{year}{2014}).

\bibitem{plaks2015}
\bibinfo{author}{Plaks, V.}, \bibinfo{author}{Kong, N.} \&
  \bibinfo{author}{Werb, Z.}
\newblock \bibinfo{title}{The cancer stem cell niche: how essential is the
  niche in regulating stemness of tumor cells?}
\newblock \emph{\bibinfo{journal}{Cell stem cell}}
  \textbf{\bibinfo{volume}{16}}, \bibinfo{pages}{225--238}
  (\bibinfo{year}{2015}).

\bibitem{messer2017}
\bibinfo{author}{Messer, J.~S.}, \bibinfo{author}{Liechty, E.~R.},
  \bibinfo{author}{Vogel, O.~A.} \& \bibinfo{author}{Chang, E.~B.}
\newblock \bibinfo{title}{Evolutionary and ecological forces that shape the
  bacterial communities of the human gut}.
\newblock \emph{\bibinfo{journal}{Mucosal immunology}}
  \textbf{\bibinfo{volume}{10}}, \bibinfo{pages}{567} (\bibinfo{year}{2017}).

\bibitem{berger2017}
\bibinfo{author}{Berger, C.~N.} \emph{et~al.}
\newblock \bibinfo{title}{Citrobacter rodentium subverts atp flux and
  cholesterol homeostasis in intestinal epithelial cells in vivo}.
\newblock \emph{\bibinfo{journal}{Cell metabolism}}
  \textbf{\bibinfo{volume}{26}}, \bibinfo{pages}{738--752}
  (\bibinfo{year}{2017}).

\bibitem{Stocker11}
\bibinfo{author}{Stocker, R.}
\newblock \bibinfo{title}{Reverse and flick: Hybrid locomotion in bacteria}.
\newblock \emph{\bibinfo{journal}{Proc Natl Acad Sci U S A}}
  \textbf{\bibinfo{volume}{108}}, \bibinfo{pages}{2635--2636}
  (\bibinfo{year}{2011}).

\bibitem{Turner00}
\bibinfo{author}{Turner, L.}, \bibinfo{author}{Ryu, W.~S.} \&
  \bibinfo{author}{Berg, H.~C.}
\newblock \bibinfo{title}{Real-time imaging of fluorescent flagellar
  filaments}.
\newblock \emph{\bibinfo{journal}{Journal of bacteriology}}
  \textbf{\bibinfo{volume}{182}}, \bibinfo{pages}{2793--2801}
  (\bibinfo{year}{2000}).

\bibitem{Berg07torque}
\bibinfo{author}{Darnton, N.~C.}, \bibinfo{author}{Turner, L.},
  \bibinfo{author}{Rojevsky, S.} \& \bibinfo{author}{Berg, H.~C.}
\newblock \bibinfo{title}{On torque and tumbling in swimming escherichia coli}.
\newblock \emph{\bibinfo{journal}{J Bacteriol}} \textbf{\bibinfo{volume}{189}},
  \bibinfo{pages}{1756--1764} (\bibinfo{year}{2007}).

\bibitem{SourjikArmitageRev}
\bibinfo{author}{Sourjik, V.} \& \bibinfo{author}{Armitage, J.~P.}
\newblock \bibinfo{title}{Spatial organization in bacterial chemotaxis}.
\newblock \emph{\bibinfo{journal}{EMBO J}} \textbf{\bibinfo{volume}{29}},
  \bibinfo{pages}{2724--2733} (\bibinfo{year}{2010}).

\bibitem{PortWadArm2008}
\bibinfo{author}{Porter, S.~L.}, \bibinfo{author}{Wadhams, G.~H.} \&
  \bibinfo{author}{Armitage, J.~P.}
\newblock \bibinfo{title}{\emph{Rhodobacter sphaeroides}: complexity in
  chemotactic signalling}.
\newblock \emph{\bibinfo{journal}{Trends in Microbiology}}
  \textbf{\bibinfo{volume}{16}}, \bibinfo{pages}{251 -- 260}
  (\bibinfo{year}{2008}).

\bibitem{Porter11}
\bibinfo{author}{Porter, S.~L.}, \bibinfo{author}{Wadhams, G.~H.} \&
  \bibinfo{author}{Armitage, J.~P.}
\newblock \bibinfo{title}{Signal processing in complex chemotaxis pathways}.
\newblock \emph{\bibinfo{journal}{Nature Reviews Microbiology}}
  \textbf{\bibinfo{volume}{9}}, \bibinfo{pages}{153--165}
  (\bibinfo{year}{2011}).

\bibitem{TuShimBerg08}
\bibinfo{author}{Tu, Y.}, \bibinfo{author}{Shimizu, T.~S.} \&
  \bibinfo{author}{Berg, H.~C.}
\newblock \bibinfo{title}{Modeling the chemotactic response of
  \emph{Escherichia coli} to time-varying stimuli.}
\newblock \emph{\bibinfo{journal}{Proc Natl Acad Sci U S A}}
  \textbf{\bibinfo{volume}{105}}, \bibinfo{pages}{14855--14860}
  (\bibinfo{year}{2008}).

\bibitem{ShimTuBerg10}
\bibinfo{author}{Shimizu, T.~S.}, \bibinfo{author}{Tu, Y.} \&
  \bibinfo{author}{Berg, H.~C.}
\newblock \bibinfo{title}{A modular gradient-sensing network for chemotaxis in
  \emph{Escherichia coli} revealed by responses to time-varying stimuli.}
\newblock \emph{\bibinfo{journal}{Mol Syst Biol}} \textbf{\bibinfo{volume}{6}},
  \bibinfo{pages}{382} (\bibinfo{year}{2010}).

\bibitem{MehGoyWin2009}
\bibinfo{author}{Mehta, P.}, \bibinfo{author}{Goyal, S.},
  \bibinfo{author}{Long, T.}, \bibinfo{author}{Bassler, B.~L.} \&
  \bibinfo{author}{Wingreen, N.~S.}
\newblock \bibinfo{title}{Information processing and signal integration in
  bacterial quorum sensing.}
\newblock \emph{\bibinfo{journal}{Mol Syst Biol}} \textbf{\bibinfo{volume}{5}},
  \bibinfo{pages}{325} (\bibinfo{year}{2009}).

\bibitem{taillefumier2015}
\bibinfo{author}{Taillefumier, T.} \& \bibinfo{author}{Wingreen, N.~S.}
\newblock \bibinfo{title}{Optimal census by quorum sensing}.
\newblock \emph{\bibinfo{journal}{PLoS computational biology}}
  \textbf{\bibinfo{volume}{11}}, \bibinfo{pages}{e1004238}
  (\bibinfo{year}{2015}).

\bibitem{DePalo17}
\bibinfo{author}{De~Palo, G.}, \bibinfo{author}{Yi, D.} \&
  \bibinfo{author}{Endres, R.~G.}
\newblock \bibinfo{title}{A critical-like collective state leads to long-range
  cell communication in dictyostelium discoideum aggregation}.
\newblock \emph{\bibinfo{journal}{PLoS Biol}} \textbf{\bibinfo{volume}{15}},
  \bibinfo{pages}{e1002602} (\bibinfo{year}{2017}).

\bibitem{Tweedy16}
\bibinfo{author}{Tweedy, L.}, \bibinfo{author}{Knecht, D.~A.},
  \bibinfo{author}{Mackay, G.~M.} \& \bibinfo{author}{Insall, R.~H.}
\newblock \bibinfo{title}{Self-generated chemoattractant gradients: attractant
  depletion extends the range and robustness of chemotaxis}.
\newblock \emph{\bibinfo{journal}{PLoS Biol}} \textbf{\bibinfo{volume}{14}},
  \bibinfo{pages}{e1002404} (\bibinfo{year}{2016}).

\bibitem{ColinSourjik17}
\bibinfo{author}{Colin, R.}, \bibinfo{author}{Rosazza, C.},
  \bibinfo{author}{Vaknin, A.} \& \bibinfo{author}{Sourjik, V.}
\newblock \bibinfo{title}{Multiple sources of slow activity fluctuations in a
  bacterial chemosensory network}.
\newblock \emph{\bibinfo{journal}{eLife}} \textbf{\bibinfo{volume}{6}},
  \bibinfo{pages}{e26796} (\bibinfo{year}{2017}).

\bibitem{Keegstra2017}
\bibinfo{author}{Keegstra, J.~M.} \emph{et~al.}
\newblock \bibinfo{title}{Phenotypic diversity and temporal variability in a
  bacterial signaling network revealed by single-cell fret}.
\newblock \emph{\bibinfo{journal}{eLife}} \textbf{\bibinfo{volume}{6}},
  \bibinfo{pages}{e27455} (\bibinfo{year}{2017}).

\bibitem{EmonetFrankel_eLife15}
\bibinfo{author}{Frankel, N.~W.} \emph{et~al.}
\newblock \bibinfo{title}{Adaptability of non-genetic diversity in bacterial
  chemotaxis}.
\newblock \emph{\bibinfo{journal}{eLife}} \textbf{\bibinfo{volume}{3}},
  \bibinfo{pages}{e03526} (\bibinfo{year}{2014}).

\end{thebibliography}

\end{document}